\documentclass[a4paper,12pt]{article}

\usepackage{amsmath,amsfonts,amssymb,mathtools}
\usepackage{mathbbol,mathrsfs,dsfont,bm}
\usepackage{latexsym,wasysym}
\usepackage{cancel,slashed,units}

\usepackage{hepnicenames}

\usepackage{graphicx,rotating}
\usepackage[table,xcdraw,dvipsnames]{xcolor}

\usepackage{booktabs,multirow,multicol}
\usepackage{enumerate,expdlist}

\usepackage{soul,comment,xstring}
\usepackage[normalem]{ulem}

\usepackage[nosort]{cite}
\usepackage{hyperref}
\usepackage{tikz}\usetikzlibrary{arrows.meta}

\hypersetup{
  colorlinks=true,
  bookmarksopen=true,
  bookmarksnumbered=true,
  pdfstartview=FitH,
  linkcolor=blue!60!black,
  urlcolor=red!60!black,
  citecolor=green!50!black
}

\allowdisplaybreaks

\newcommand{\ran}{\textrm{--}}

\newcommand{\mio}[1]{}
\newcommand{\gfootnote}[1]{}

\newcommand{\beq}{\begin{equation}}
\newcommand{\eeq}{\end{equation}}
\newcommand{\be}{\begin{equation}}
\newcommand{\ee}{\end{equation}}

\newcommand{\beqa}{\begin{equation}\begin{aligned}}
\newcommand{\eeqa}{\end{aligned}\end{equation}}

\newcommand{\bea}{\begin{eqnarray}}
\newcommand{\eea}{\end{eqnarray}}

\newcommand{\bpm}{\begin{pmatrix}}
\newcommand{\epm}{\end{pmatrix}}

\renewcommand{\[}{\left[}

\newcommand{\Lag}{\mathscr{L}}

\newcommand{\One}{\mathbb{1}}

\newcommand{\med}[1]{\langle #1\rangle}
\newcommand{\sfrac}[2]{#1/#2}

\renewcommand{\Im}{\operatorname{Im}}
\renewcommand{\Re}{\operatorname{Re}}
\newcommand{\Tr}{\,\operatorname{Tr}}

\newcommand{\SU}{\,\mathrm{SU}}

\newcommand{\U}{\,\mathrm{U}}

\newcommand{\bp}{\bar{M}_{\mathrm{Pl}}}

\newcommand{\fig}[1]{~\ref{fig:#1}}
\newcommand{\eq}[1]{~\eqref{eq:#1}}

\newcommand{\hhref}[1]{%
  \href{https://arxiv.org/abs/#1}{arXiv:#1}%
}
\newcommand{\hhrefq}[1]{%
  \IfSubStr{#1}{:}
    {\href{https://inspirehep.net/search?q=#1}{InSpire:#1}}
    {\hhref{#1}}%
}
\newcommand{\hhrefqq}[1]{%
  \IfBeginWith{#1}{10.}
    {\href{https://doi.org/#1}{doi:#1}}
    {\hhrefq{#1}}%
}

\newcommand{\doi}[1]{\href{https://doi.org/#1}{[link]}}

\makeatletter

\def\art{\@ifnextchar[{\eart}{\oart}}
\def\oart#1#2#3#4#5{%
  \textrm{#1}, \textit{#2 \textbf{#3}} \textrm{(#5) #4}%
}
\def\eart[#1]#2#3#4#5#6{%
  \oart{#2}{#3}{#4}{#5}{#6} (\textit{#1})%
}

\def\article{\@ifnextchar[{\earticle}{\oarticle}}
\def\oarticle#1#2#3#4#5#6{%
  \textrm{#1}, \textit{`#6'}, \textrm{#2 #3 (#5) #4}%
}
\def\earticle[#1]#2#3#4#5#6#7{%
  \oarticle{#2}{#3}{#4}{#5}{#6}{#7} [\hhrefq{#1}]%
}

\def\hepart[#1]#2{\textrm{#2}, {\slshape #1}}
\def\heparticle[#1]#2#3{%
  #2, \textit{`#3'} [\hhrefq{#1}]%
}

\renewenvironment{thebibliography}[1]
  {%
    \begin{multicols}{2}[\section*{\refname}]%
    \@mkboth{\MakeUppercase\refname}{\MakeUppercase\refname}%
    \list{\@biblabel{\@arabic\c@enumiv}}%
      {%
        \settowidth\labelwidth{\@biblabel{#1}}%
        \leftmargin\labelwidth
        \advance\leftmargin\labelsep
        \@openbib@code
        \usecounter{enumiv}%
        \let\p@enumiv\@empty
        \renewcommand\theenumiv{\@arabic\c@enumiv}%
      }%
    \sloppy
    \clubpenalty4000
    \@clubpenalty\clubpenalty
    \widowpenalty4000
    \sfcode`\.\@m
  }
  {%
    \def\@noitemerr{%
      \@latex@warning{Empty `thebibliography' environment}%
    }%
    \endlist
    \end{multicols}%
  }

\makeatother

\newcounter{alphaequation}[equation]
\renewcommand{\thealphaequation}{%
  \theequation\hbox to 0.6em{\hfil\alph{alphaequation}\hfil}%
}

\makeatletter

\newcommand{\eqnsystem}[1]{%
  \renewcommand{\@eqnnum}{\textrm{(\thealphaequation)}}%
  \renewcommand{\@@eqncr}{%
    \let\@tempa\relax
    \ifcase\@eqcnt
      \def\@tempa{& & &}%
    \or
      \def\@tempa{& &}%
    \or
      \def\@tempa{&}%
    \fi
    \@tempa
    \if@eqnsw
      \@eqnnum\refstepcounter{alphaequation}%
    \fi
    \global\@eqnswtrue
    \global\@eqcnt=0
    \cr
  }%
  \refstepcounter{equation}%
  \let\@currentlabel\theequation
  \def\@tempb{#1}%
  \ifx\@tempb\empty\else\label{#1}\fi
  \refstepcounter{alphaequation}%
  \let\@currentlabel\thealphaequation
  \global\@eqnswtrue
  \global\@eqcnt=0
  \tabskip\@centering
  \let\\=\@eqncr
  $$\halign to \displaywidth\bgroup
    \@eqnsel\hskip\@centering
    $\displaystyle\tabskip\z@{##}$%
    &\global\@eqcnt\@ne
    \hskip2\arraycolsep\hfil${##}$\hfil
    &\global\@eqcnt\tw@
    \hskip2\arraycolsep
    $\displaystyle\tabskip\z@{##}$\hfil
    \tabskip\@centering
    &\llap{##}\tabskip\z@\cr
}

\def\endeqnsystem{%
  \@@eqncr\egroup$$\global\@ignoretrue
}

\makeatother

\begin{document}

\begin{center}
  \bigskip\bigskip
  {\bfseries\LARGE\color{red!40!black} Ohmic warm inflation}\\
  \bigskip\bigskip
  {\bfseries Anish Ghoshal$^a$, Alessandro Strumia$^b$}\\[2ex]
  {\itshape $^a$Department of Physics and Astronomy, University of Sussex, UK}\\
  {\itshape $^b$Dipartimento di Fisica, Universit\`a di Pisa, Italia}\\
  \bigskip
  {\bfseries\large\color{blue!30!black} Abstract}
  \begin{quote}\large\color{blue!30!black} 
We show that Warm Little Inflation is anomalous, lacks a suitable inflaton potential,
and cannot sustain the thermal bath, as the quasi-static approximation over-estimated the dissipated power.
To address these issues, we propose a new friction mechanism and new anomaly-free models
based on one fermion charged under a U(1) gauge group in one extra dimension.
The four-dimensional lattice version is a ring of $N$ U(1) groups with one fermion at each site, reducing to an anomaly-free
version of the warm little inflaton model for $N=2$.
The inflaton is a Wilson line, protected for $N\ge5$.
Large $N$ allows a suitable monodromy inflaton potential.
In the continuum limit, $N\to\infty$, the inflaton velocity becomes a five-dimensional electric field
and friction becomes Ohmic dissipation, 
which seems able to sustain warm inflation
compatibly with observations.
  \end{quote}
\end{center}
\thispagestyle{empty}

\tableofcontents

\section{Introduction}\label{sec:intro}
Warm inflation~\cite{Berera,astro-ph/9501024} assumes that the inflaton $\phi$ dissipates energy into a thermal bath during slow roll, through a friction coefficient $\Upsilon$ large enough to sustain a temperature $T>H$ despite the expansion~\cite{0808.1855,2302.02827}.
The main obstacle to a realisation in quantum field theory is that particles coupled to $\phi$ strongly enough to give efficient friction 
also induce thermal corrections to the inflaton mass that spoil slow roll~\cite{hep-ph/9803394,hep-ph/9809409}.
Couplings to heavy mediators avoid the thermal corrections, but need large multiplicities~\cite{1008.1929,1207.0445}.

\smallskip

The `warm little inflaton' (WLI) model~\cite{1604.08838} addresses both issues through collective symmetry breaking, as in little-Higgs models~\cite{hep-ph/0105239,hep-ph/0502182}.
The inflaton is the relative phase $\theta$ of two scalars that give masses $m_1\propto\cos(\theta/2)$ and $m_2\propto\sin(\theta/2)$ to two fermions.
These masses depend on the inflaton, while the leading thermal correction to the potential, proportional to $m_1^2+m_2^2$, does not depend on $\theta$.
The fermions are kept close to equilibrium by Yukawa couplings to a bath, and produce a friction $\Upsilon\propto T$.
Detailed studies found consistent inflationary trajectories in agreement with cosmic microwave background (CMB) data, even for a quartic potential~\cite{1604.08838,1805.07186,1710.10008,2504.02655}.
The model has been embedded in extensions of the Standard Model with right-handed neutrinos~\cite{2012.03988}, connected to dark matter, dark energy~\cite{1811.05493,1906.11835} and the QCD axion~\cite{2105.05771,2609.19284}, modified with a scalar bath that allows strong dissipation~\cite{1907.13410}, and reduced to a single fermion~\cite{2308.00564}.

Section~\ref{sec:WLI} describes the original model and tests whether the dynamics assumed in these analyses is consistent with the underlying theory.
Problems with gauge anomalies and thermal masses can be fixed (section~\ref{sec:fix});
but the friction was over-estimated by assuming that the fermion occupations follow their oscillating masses: 
even optimal relaxation supplies at most about 10\% of the heating needed to sustain thermal inflation (section~\ref{sec:friction}).
Pair production of fermions provides a new friction mechanism (section~\ref{sec:fixfriction})
and allows a good ultraviolet behaviour of the couplings (section~\ref{sec:UV}).
However, the periodicity of the inflaton clashes with the needed inflationary potential (section~\ref{sec:angle}).

In section~\ref{sec:ring} we propose new anomaly-free models with fermions on a ring of $N$ U(1) gauge groups,
the lattice version of a fermion charged under a U(1) in one compact extra dimension.
The inflaton is a Wilson line whose potential is protected for $N\ge5$.
An acceptable inflaton potential arises when the U(1) is broken by a charged scalar (sections~\ref{sec:5dmono} and~\ref{sec:JJ}).
Section~\ref{sec:ringwarm} studies warm inflation in these models at finite $N$.
Section~\ref{sec:WI5d} studies them in the continuum limit, where inflaton motion becomes an electric field along the extra dimension
and friction becomes Ohmic dissipation. Conclusions are given in section~\ref{sec:concl}.

Our main results are: occupation relaxation is limited by the frequency at which fermion masses are modulated, rather than by the Hubble rate;
locality in theory space protects the inflaton potential of the ring models;
in the continuum limit, dissipation becomes Ohmic conduction along the compact dimension.

\section{Warm little inflation and its issues}\label{sec:WLI}
Friction requires bath particles coupled strongly enough to the inflaton, but the same couplings correct the inflaton potential and can spoil its flatness.
Fermions offer a possible way to separate these effects:
their thermal corrections depend on the inflaton only through traces of the mass matrix, while dissipation depends on the individual masses.
Indeed all fermionic Matsubara frequencies are at least $\pi T$,
so that for $T\gg M$ the potential is analytic in the Dirac mass matrix $M$.\footnote{Bosons instead have a zero Matsubara mode which, if unscreened, generates the non-analytic term $-T\Tr(M^2)^{3/2}/12\pi$, not protected by the cancellation of a few polynomial traces.}
The $\ln M^2$ terms present separately in the vacuum and thermal parts cancel in their sum. Expanding gives the one-loop correction
\beq
\delta V= T^4 \sum_{n=0}^\infty c_n\, \Tr\bigg(\frac{M^\dagger M}{\pi^2T^2}\bigg)^n,\qquad
c_n= \frac{(-1)^n}{{2\,n!}} {\pi^{3/2}\,\Gamma(n-\frac32)\,(1-2^{3-2n})\,\zeta(2n-3)} 
\label{eq:Fanalytic}
\eeq
valid for all $n\neq2$, 
while the pole of $\zeta(1)$ becomes $c_2=\pi^2\ln(\bar\mu e^{\gamma_E}/\pi T)/8$ in the $\overline{\rm MS}$ scheme, corresponding to couplings renormalised at $\bar\mu\approx T$.
The series converges for masses below $\pi T$.

The inflaton dependence of the potential is therefore controlled by the traces, and a symmetry that makes $\Tr[(M^\dagger M)^n]$ inflaton-independent up to large enough $n$ protects it to the corresponding order in $(M/T)^2$, while leaving the individual masses inflaton dependent and available to source dissipation.
The models discussed in this paper follow this logic.

\medskip

The ``warm little inflaton'' model~\cite{1604.08838} employs a gauged $\U(1)_X$ with gauge coupling $g_X$, two complex scalars $\Phi_{1,2}$ with equal unit charge, two Dirac fermions $\psi_{1,2}$ whose left-handed components have unit charge, and whose right-handed components are neutral.
Furthermore, a thermal bath is modelled as a neutral scalar $\sigma$ and a Dirac fermion $\psi_\sigma$ with unit charge for $\psi_{\sigma R}$.
Besides gauge invariance, the model assumes the interchange symmetry $\Phi_1\leftrightarrow i\Phi_2$, $\psi_1\leftrightarrow\psi_2$, which equates the Yukawa couplings of the two fermions~\cite{1604.08838}.
The scalar potential and the  Yukawa interactions are assumed to be
\begin{eqnarray}
V &=& -m_\Phi^2(|\Phi_1|^2+|\Phi_2|^2)+\lambda(|\Phi_1|^4+|\Phi_2|^4)+\lambda_{12}|\Phi_1|^2|\Phi_2|^2,\\
\Lag_{\rm Y}&=&-\frac{y}{2}\left[(\Phi_1+\Phi_2)\,\bar\psi_{1L}\psi_{1R}-i(\Phi_1-\Phi_2)\,\bar\psi_{2L}\psi_{2R}\right]-{y'}\,\sigma\sum_{i=1,2}(\bar\psi_{iL}\psi_{\sigma R}+\bar\psi_{\sigma L}\psi_{iR})+{\rm h.c.}~
\label{eq:Lorig}
\end{eqnarray}
The potential is invariant under separate rephasing of $\Phi_1$ and $\Phi_2$, so it is flat along $\Phi_{1,2}=fe^{\pm i\theta/2}$. 
The relative phase $\theta$ is the inflaton, with an extra inflationary potential added separately.
However the assumed symmetries  allow extra potential terms involving $\Phi_1^*\Phi_2$
which give a potential to $\theta$: they cannot be forbidden, since the Yukawa couplings  break the relative rephasing, and must be assumed small.
The scalar kinetic terms along the flat direction are
\be
\Lag_{\rm kin}=2(\partial_\mu f)^2+\frac{f^2}2(\partial_\mu\theta)^2,
\label{eq:Lkin}
\ee
so that at fixed radius the canonical inflaton is $\phi=f\theta$, periodic with period $2\pi f$.
The vector eats the common phase and acquires mass $M_X=2g_Xf$.
The masses of the $\psi_{1,2}$ fermions  are\footnote{The model is simpler in terms of $S_1=(\Phi_1+\Phi_2)/\sqrt2$ and $S_2=-i(\Phi_1-\Phi_2)/\sqrt2$, which also have unit charge: each fermion couples to one scalar, $y\,S_i\bar\psi_{iL}\psi_{iR}/\sqrt2$, and the theory is symmetric under $S_1\leftrightarrow S_2$.
The flat direction becomes $S_1=\sqrt2f\cos(\theta/2)$, $S_2=\sqrt2f\sin(\theta/2)$, and the potential above becomes eq.\eq{V0} with $\lambda_1=\lambda+\lambda_{12}/2$, $\lambda_2=2\lambda-\lambda_{12}$ and $\lambda_5=\lambda_6=0$.
In~\cite{1604.08838,1805.07186} the Yukawa couplings are denoted $g$ and $h={y'}$, and $\Phi_{1,2}=Me^{\pm i\phi/M}/\sqrt2$, so that $f=M/\sqrt2$, $y=\sqrt2g$, and the canonical inflaton is $f\theta=\sqrt2\phi$.
Their numerical analyses treat $\phi$ as canonical: this changes $y$ and $T/f$ by factors $\sqrt2$, which do not affect our conclusions.}
\be
 m_1=yf\cos\frac\theta2,\qquad m_2=yf\sin\frac\theta2,
\label{eq:masses}		
\ee
The inflaton dependence of individual fermion masses will provide friction,
while the inflaton potential does not receive quadratically divergent correction nor a thermal term $T^2\sum_i|m_i|^2/12$
as $\sum_i|m_i|^2=y^2f^2$ is independent of $\theta$.
The quartic combination $\sum_i|m_i|^4$ depends on $\theta$, so the
 resulting $\delta V \sim y^4 f^4 \cos(2\theta)/(4\pi)^2$ needs to be small enough to not spoil warm inflation.


\medskip

The homogeneous inflaton and the radiation energy density $\rho_R=\pi^2g_*T^4/30$ evolve as
\be
\ddot\phi+(3H+\Upsilon)\dot\phi+V'(\phi)=0,\qquad \dot\rho_R+4H\rho_R=\Upsilon\dot\phi^2,
\label{eq:eom}
\ee
where $g_*$ counts the relativistic degrees of freedom. 
The relevance of the extra friction term $\Upsilon$ is quantified by  $Q\equiv\Upsilon/3H$.
According to~\cite{1604.08838} (see also~\cite{2210.08762})  the dissipation coefficient is estimated as $\Upsilon\approx C_TT$ with
\be
C_T\approx\frac{3y^2}{4{y'}^2(1-0.34\ln {y'})}.
\label{eq:CT}
\ee
Section~\ref{sec:friction} will critically review this result.
The curvature power spectrum, measured as ${\cal P}_R\approx2.1~10^{-9}$, is~\cite{1302.3544,1401.1149}
\be
{\cal P}_R=G(Q)\,(1+R)\left(\frac{H^2}{2\pi\dot\phi}\right)^2,\qquad R=2n_*+\frac{2\pi Q}{\sqrt{1+4\pi Q/3}}\,\frac TH,
\label{eq:PR}
\ee
where $n_*$ is the occupation number of inflaton modes at horizon exit.
The fit of~\cite{1805.07186} assumes thermalised inflaton modes, $n_*\approx T/H$, which enhances the power spectrum by a factor $\sim2T/H$; 
however $\Upsilon<H$ for $Q<1/3$ cannot thermalise horizon-scale modes within a Hubble time, so vacuum occupation $n_*=0$ is equally possible.
The factor $G$ accounts for the coupling between inflaton and radiation perturbations: for $\Upsilon\propto T^c$ with $c>0$ it grows for $Q\gtrsim1$, so that strong dissipation at horizon exit is disfavoured by data~\cite{0905.3500,1106.0701}.

The analysis of~\cite{1805.07186} imposed light fermions and broken symmetry, $yf\lesssim T\lesssim \sqrt2f$, fermions close to equilibrium, with decay widths $\Gamma_{\psi_i}$ into the bath larger than the Hubble rate, $\Gamma_{\psi_i}>H$, and $T>H$.
Assuming a quartic inflaton potential it found consistent solutions for $y\approx0.014$-$0.14$, ${y'}\approx2$-$3$ and $Q_*\lesssim0.2$ at horizon exit, with $Q$ growing until inflation ends in the strong regime.
In the stationary regime $\rho_R=\Upsilon\dot\phi^2/4H=3Q\dot\phi^2/4$, so $\dot\phi^2=2\pi^2g_*T^4/45Q$, and the angular velocity at a generic radius, with $\dot\phi\equiv f\dot\theta$ the tangential speed, is
\be
\frac{\dot\theta}{T}=\frac{T}{f}\sqrt{\frac{2\pi^2g_*}{45\,Q}}.
\label{eq:deltadot}
\ee
For weak dissipation the square root is of order a few, so that $\dot\theta\sim T$ for $T\sim f$, as in the benchmark of~\cite{1604.08838}, where the fermion masses oscillate with angular frequency $\approx10^3H$~\cite{2308.00564}.

In the rest of this section we list different issues and address them.

\subsection{Issue 1: gauge anomalies}\label{sec:anomaly}
With this field content, the chiral gauge symmetry has uncancelled anomalies: $\sum q_X=\sum q_X^3=1$ so the $\U(1)_X^3$ and the mixed gravitational anomalies are non-zero.
Anomalies are not discussed in~\cite{1604.08838,1805.07186,1811.05493,2308.00564}.
The model of~\cite{2012.03988}, where $\psi_{1,2}$ are right-handed neutrinos happens to be anomaly free for $\U(1)_X=B-L$ with the standard charges of quarks and leptons, while its alternative identification with lepton number is anomalous.
These anomalies can be cancelled by modifying the fermion charges, as described below, where we also address the second issue.

\subsection{Issue 2: thermal masses}\label{sec:thmass}
Since both fermions couple to the same bath state, the thermal self-energy is a matrix in flavour space.
At leading order it is ${y'}^2T^2J/8$, where $J$ is the $2\times2$ matrix with all entries equal to one.
Its eigenvalues give the asymptotic squared mass ${y'}^2T^2/4$ to $\psi_+=(\psi_1+\psi_2)/\sqrt2$, and no contribution to $\psi_-=(\psi_1-\psi_2)/\sqrt2$, rather than ${y'}^2T^2/8$ to each fermion as assumed in~\cite{1604.08838,1805.07186}.
Keeping $\psi_+$ lighter than $T$ requires roughly ${y'}<2$, at the edge of the values ${y'}\approx2$-$3$ found in~\cite{1805.07186}.
%

Moreover, only $\psi_+$ couples to the bath, while the inflaton-dependent masses mix it with $\psi_-$: the friction must then be computed for a $2\times2$ density matrix of occupations and coherences, rather than for independent occupations.
The same issue arises in the embedding of~\cite{2012.03988}, where the two right-handed neutrinos couple to the same lepton-Higgs combination.

The Yukawa couplings $y$ respect the structure that cancels the $T^2$ term, but the common-bath coupling ${y'}$ does not: it allows other combinations of $\Phi_{1,2}$ in the Yukawa couplings, and potential terms such as $T^2\Im(\Phi_1^*\Phi_2)$.
Their size needs a dedicated computation: a single insertion of the common-bath self-energy gives $\Tr(JM^\dagger M)=|m_1|^2+|m_2|^2$, still independent of $\theta$.

%

\subsection{Fixing issues 1 and 2: an anomaly-free model}\label{sec:fix}
Both issues are avoided by giving the two fermions opposite chiral charges, and a separate bath to each of them, described as a neutral Dirac fermion $b_i$ and a complex scalar $\chi_i$.
The assumed charges are:
\be
\begin{array}{c|cccccccc}
\text{field} & S_{1,2} & \psi_{1L} & \psi_{1R} & \psi_{2L} & \psi_{2R} & b_{iL,R} & \chi_1 & \chi_2\\
\hline
q_X & 1 & \tfrac12 & -\tfrac12 & -\tfrac12 & \tfrac12 & 0 & \tfrac12 & -\tfrac12
\end{array}.
\ee
The cubic and mixed gravitational anomalies cancel between the two active fermions~\cite{1905.13729}.
The Yukawa interactions are
\be\label{eq:Lpsi}
-\Lag_{\rm Yuk}=\frac{y}{\sqrt2}(S_1\bar\psi_{1L}\psi_{1R}+S_2^*\bar\psi_{2L}\psi_{2R})+
{y'}\sum_{i=1,2}\left(\chi_i\,\bar\psi_{iL}b_{iR}+\chi_i^*\,\bar\psi_{iR}b_{iL}\right)+{\rm h.c.}
\ee
Equal couplings are enforced by assuming the exchange symmetry
\be
S_1\leftrightarrow S_2^*,\qquad \psi_1\leftrightarrow\psi_2,\qquad b_1\leftrightarrow b_2,\qquad \chi_1\leftrightarrow\chi_2,\qquad X_\mu\to-X_\mu.
\ee
Gauge and exchange symmetries still allow crossed couplings such as $S_2\bar\psi_{1L}\psi_{1R}$, as well as Dirac masses mixing $\psi_1$ and $\psi_2$.
They are forbidden by a $\mathbb{Z}_4^{(1)}\times\mathbb{Z}_4^{(2)}$ symmetry, where $\mathbb{Z}_4^{(1)}$ acts as
\be
S_1\to-S_1,\qquad \psi_{1L}\to i\psi_{1L},\qquad \psi_{1R}\to-i\psi_{1R},\qquad \chi_1\to i\chi_1,
\ee
and $\mathbb{Z}_4^{(2)}$ is its image under the exchange.
These symmetries still allow the bath cubic $A_\chi(S_1^*\chi_1^2+S_2\chi_2^2)+{\rm h.c.}$, 
and Dirac masses for the neutral $b_i$.
Both are forbidden by a further $\U(1)$ for each family, under which $\chi_i$ and $b_{iL}$ have charge $+1$ and $b_{iR}$ charge $-1$, interchanged by the exchange symmetry.
It allows the portal couplings $|\chi_i|^2|S_j|^2$, which we assume to be small.
Up to an additive constant, the most general renormalisable potential involving only $S_{1,2}$ and invariant under these symmetries is
\be\begin{aligned}
V_0=&-m_\Phi^2(|S_1|^2+|S_2|^2)+\frac{\lambda_1}{2}(|S_1|^2+|S_2|^2)^2+\lambda_2[\Im(S_1^*S_2)]^2\\
&+\frac{\lambda_5}{2}\left[(|S_1|^2-|S_2|^2)^2-4[\Re(S_1^*S_2)]^2\right]+\lambda_6\Re(S_1^*S_2)\Im(S_1^*S_2),
\end{aligned}\label{eq:V0}\ee
with real coefficients.
For $\lambda_5=\lambda_6=0$, $m_\Phi^2>0$, $\lambda_1>0$ and $\lambda_2>0$, the potential favours aligned phases and
becomes invariant under continuous real rotations of $(S_1,S_2)$, so that its minimum has a continuous family of degenerate vacua
\be
\langle S_1\rangle=\sqrt2f\cos\frac\theta2,\qquad
\langle S_2\rangle=\sqrt2f\sin\frac\theta2,\qquad f^2=\frac{m_\Phi^2}{2\lambda_1}.
\ee
Along the flat direction the potential reduces to
\be
V_0=-2m_\Phi^2f^2+2\lambda_1f^4+2\lambda_5f^4\cos2\theta.
\ee
The CP-odd $\lambda_6$ term vanishes on this trajectory, but minimising over $\Im(S_1^*S_2)$ gives at leading order $-\lambda_6^2[\Re(S_1^*S_2)]^2/4\lambda_2\propto\cos^2\theta$, which adds to the $\cos2\theta$ term.
Thus the light angular direction requires small $\lambda_5$ and $\lambda_6$, which is not enforced by symmetry.
Below we set $\lambda_{5,6}$ and the portal couplings to zero at the matching scale $\bar\mu\sim f$.
At leading order in ${y'}$, $\psi_i$ and $b_i$ acquire the asymptotic thermal mass $m_T^2={y'}^2T^2/8$, and $\chi_i$ acquires ${y'}^2T^2/6$, plus contributions from its self-coupling and from $g_X$.
The model has $g_*=18$ degrees of freedom counting $\psi_i$, $b_i$ and $\chi_i$ as massless, and excluding the gauge and Higgs excitations.

\subsection{Issue 3: centrifugal displacement}\label{sec:rotation}
On the assumed trajectory the kinetic energy is $K=2\dot f^2+f^2\dot\theta^2/2$, and the thermal free energy is $\approx-2m_\Phi^2(T)f^2+2\lambda_1f^4$ with
\be
m_\Phi^2(T)=m_\Phi^2-\left(\frac{\lambda_1}{4}+\frac{\lambda_2}{24}+\frac{g_X^2}4+\frac{y^2}{24}\right)T^2.
\ee
The rotation pushes the condensates outwards:
neglecting radial acceleration and friction, the radial forces balance when
\be
2\lambda_1f^2=m_\Phi^2(T)+\frac{\dot\theta^2}4.
\label{eq:radial}
\ee
The static radius $f_{\rm mf}^2=m_\Phi^2(T)/2\lambda_1$ applies only if $\dot\theta$ is small compared to the radial mass. 
At fixed canonical speed $\dot\phi$ one finds
\be
f^2=\frac{m_\Phi^2(T)+\sqrt{m_\Phi^4(T)+2\lambda_1\dot\phi^2}}{4\lambda_1},\qquad
\eta_r\equiv\frac{\lambda_1\dot\phi^2}{2m_\Phi^4(T)}=\frac{\pi^2g_*}{180\,Q\,\lambda_1}\left(\frac{T}{f_{\rm mf}}\right)^4,
\label{eq:vrot}
\ee
where $\eta_r$ is the ratio between the angular kinetic energy and twice the depth of the thermal potential, and the last expression uses energy balance.
For $Q\lesssim0.2$ and quartic couplings of order unity $\eta_r\gtrsim1$ unless $T\ll f_{\rm mf}$: the radius then grows by order one, and the vector and radial modes, with masses of a few $T$, are not decoupled.
The constant $f$ assumed in~\cite{1604.08838,1805.07186} is therefore not a good approximation, e.g.\ for the benchmark trajectory of~\cite{1604.08838}.
This does not by itself exclude inflation, but it requires the radial response to be included.
%

\subsection{Issue 4: not enough friction}\label{sec:friction}
The key feature of warm inflation is the dissipation coefficient $\Upsilon$, which transfers energy from the inflaton to the bath.
Friction arises because the fermion energies $E_p \approx \sqrt{p^2+|m_i(\phi)|^2+m_T^2}$ depend on $\phi$ through the Dirac masses.
A rolling inflaton changes them, while the occupations $f_p$ relax with some rate $\gamma_p$ towards the instantaneous equilibrium,
\be
\dot f_p=-\gamma_p\left[ f_p-f_{\rm F}(E_p(t))\right],
\label{eq:relax}
\ee
where $f_{\rm F}$ is the Fermi-Dirac distribution.
The force on the inflaton, $\sum_ig_i\int d^3p\,(\partial E_p/\partial\phi)f_p/(2\pi)^3$ with $g_i=4$ states per Dirac fermion, contains the thermal potential from $f_{\rm F}$, 
plus a term from the lag $\delta f_p\equiv f_p-f_{\rm F}$.
This extra term gives a friction if eq.\eq{relax} is approximately solved assuming that relaxation is the fastest process~\cite{1604.08838}.
Then the static solution of eq.\eq{relax} is $\delta f_p\approx-\dot f_{\rm F}/\gamma_p\propto\dot\phi$. For the fermions of eq.\eq{Lorig} this gives the friction~\cite{1604.08838}
\be
\Upsilon=\frac{4}{T}\sum_i\int\frac{d^3p}{(2\pi)^3}\left(\frac{\partial m_i}{\partial\phi}\right)^2\frac{m_i^2}{E_p^2\,\gamma_p}f_{\rm F}(1-f_{\rm F})\approx C_T\,T.
\label{eq:UpsWLI}
\ee
The authors of \cite{1604.08838} identified the relaxation rate with the decay width of $\psi_i$ into the bath, $\gamma_p=\Gamma_{\psi_i}$.
Since $\Gamma_{\psi_i}\propto {y'}^2m_i^2$, the factor $m_i^2$ cancels, and $\sum_i(\partial m_i/\partial\phi)^2=y^2/4$ does not depend on $\theta$, leaving $\Upsilon\propto T$.
However, this computation involves three unjustified steps:
\begin{enumerate}[a)]
\item  {\em Dirac vs thermal masses}.
The friction is controlled by $\partial E_p/\partial\phi$, and only the Dirac masses $m_i$ depend on $\phi$, while the thermal mass $m_T$ does not.
Equivalently, $\phi$ couples to the chirality-flipping operator $\bar\psi\psi$, while the thermal mass induced by ${y'}$ is chirally symmetric.
However, the factor $m_i^2$ in the numerator of eq.\eq{UpsWLI} was identified with the full mass $m_i^2+m_T^2$, and cancelled against the width $\Gamma_{\psi_i}\propto{y'}^2(m_i^2+m_T^2)$, obtaining $\Upsilon\propto T$.
Keeping only the Dirac masses, in the regime $m_T\gg yf$ assumed in~\cite{1604.08838,1805.07186} the friction gets smaller by a factor
\be
\frac{\Upsilon}{C_TT}\approx\frac{4y^2f^2}{{y'}^2T^2}\sin^2\theta.
\label{eq:Upscorr}
\ee
This estimate assumes diagonal masses and common thermal denominators, and decreases with temperature as $1/T$, so $G(Q)$ computed for $\Upsilon\propto T$ would not apply.
When the thermal masses dominate,
a precise computation would need the thermal $\bar\psi\psi$ correlator~\cite{1008.1929}.\footnote{Proposals that use the Warm Little Inflaton for a later stage of warm inflation that enhances small-scale perturbations~\cite{2410.10996,2609.19270} (see also~\cite{2105.08045} for a scalar variant)
rely on the growth of $G(Q)$ for $\Upsilon\propto T$: with the Dirac masses $\Upsilon$ decreases with $T$, and this growth is absent.}

\item {\em The relaxation rate}.
In~\cite{1604.08838,1805.07186} the width in eq.\eq{UpsWLI} was computed as the decay rate of $\psi_i$ into the bath particles, neglecting their masses.
However, the same coupling ${y'}$ gives thermal masses of order ${y'}T$ to all particles involved, which close this decay.
Thermal masses only give an approximate description: the relaxation rate is a genuinely thermal quantity, 
dominated by $2\leftrightarrow2$ scatterings and by collinear processes~\cite{hep-ph/0209353,1012.3784}.

\item Most importantly, {\em inflaton rotation is faster than relaxation, suppressing friction}.
The authors of \cite{1604.08838,1805.07186} assumed that $\gamma_p >H$ justifies the quasi-static approximation in eq.\eq{UpsWLI}.
Instead it only holds if $\gamma_p\gg\Omega \equiv \dot\theta $.
This is important because the solutions have  $\Omega \gg H$.\footnote{The fermion dispersion relation $E_p^2=p^2+|m_i|^2+m_T^2$ is modulated with frequency $\Omega=\dot\theta$, as $|m_{1,2}|^2=y^2f^2(1\pm\cos\theta)/2$.
The zeros of the signed masses $m_{1,2}$ do not make $E_p$ singular, so that the occupations respond on the time scale $1/\Omega$, rather than $|m_i/\dot m_i|$.}
Taking into account the modulation $E_p\supset\Delta E_p\cos\Omega t$, the stationary solution of eq.\eq{relax} is
\be
\delta f_p=-\Re\left[\frac{i\Omega\,e^{i\Omega t}}{\gamma_p+i\Omega}\right]\frac{\partial f_{\rm F}}{\partial E_p}\,\Delta E_p,
\ee
which reduces to the quasi-static $\delta f_p\approx-\dot f_{\rm F}/\gamma_p$ for $\gamma_p\gg\Omega$.
The cycle-averaged power absorbed per mode, $\med{f_p\,\dot E_p}$, is
\be
\med{W_p}=W_p\,\frac{\gamma_p\Omega}{\gamma_p^2+\Omega^2},\qquad
W_p=\Omega\,\mathcal E_p,\qquad
\mathcal E_p=-\frac{\partial f_{\rm F}}{\partial E_p}\,\frac{(\Delta E_p)^2}{2}.
\label{eq:lag}
\ee
\end{enumerate}
The same suppression occurs in Drude conductors or in microwave ovens, where it is known as Debye relaxation: 
water molecules dissipate the oscillating field efficiently only up to frequencies of order their relaxation rate.
Physically, the modulation pushes the occupations out of equilibrium, where they store at most the energy $\mathcal E_p$, reached when they do not respond.
This energy is exchanged with the inflaton, with power $W_p$, and collisions dissipate only a fraction $\gamma_p\Omega/(\gamma_p^2+\Omega^2) \le 1/2$ of it,
maximal for $\gamma_p=\Omega$, and suppressed otherwise.
The quasi-static eq.\eq{UpsWLI} is recovered in the limit $\gamma_p\gg\Omega$, but  \cite{1604.08838,1805.07186} applied it for $\gamma_p\ll \Omega$ obtaining
an $(\Omega/\gamma_p)^2 \sim  10^{3\ran4}$ over-estimate of the friction.

\begin{figure}[t]
\centering
\includegraphics[width=0.65\textwidth]{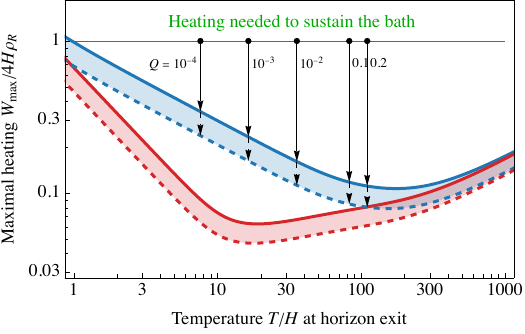}
\caption{\em Upper bound on the heating from relaxation of the fermion
occupations, divided by the heating needed to sustain the bath, for the
most favourable choices $T=f = yf$ and optimal relaxation rate
$\gamma=\Omega$, with thermalised ($n_*\approx T/H$) or vacuum ($n_*=0$)
inflaton modes at horizon exit.
Dots are the quartic solutions of~\cite{1805.07186}, and arrows show their deficit for $g_*=12.5$ (solid curves) and for $g_*=18$ of the anomaly-free model of section~\ref{sec:fix} (dashed curves).
\label{fig:WLIheating}}
\end{figure}

We now test whether even the maximal heating, obtained for $\gamma_p=\Omega$, can sustain the proposed warm inflation solutions.
Whatever the relaxation rates, the absorbed power cannot exceed half of the oscillating power\footnote{One can prove that the maximal friction
also holds for a general linearised collision term. 
Indeed, detailed balance decomposes into independently relaxing modes~\cite{hep-ph/9512263,hep-ph/0608012}, after projecting out conserved quantities, and the same argument applies to each harmonic $n\Omega$ of the drive,
with equality only if the spectral weight sits at $\gamma=n\Omega$.
Processes spread over many $n$ are subject to a stronger bound.
A response with memory obeys the bound as long as it is a superposition of decaying exponentials; 
only an oscillating response, i.e.\ a collective resonance near $n\Omega$, can exceed it.
Finally, going beyond the linear approximation, we verified that non-linear corrections are small:
a numerical solution of the full nonlinear eq.\eq{relax} exceeds the friction by less than $0.4\%$ for $yf\le2T$.
The bound does not apply to coherent flavour oscillations nor to pair production.}
\be
\med{W}\le\frac12\sum_ig_i\int\frac{d^3p}{(2\pi)^3}\,W_p=\frac{\Omega}{2}\,\mathcal E,\qquad
\mathcal E=\sum_ig_i\int\frac{d^3p}{(2\pi)^3}\,\mathcal E_p.
\label{eq:Wbound}
\ee
We next show that the maximal friction $\Upsilon=\med{W}/\dot\phi^2 \le\mathcal E/2f\dot\phi$
cannot sustain the regime assumed in~\cite{1604.08838,1805.07186}: a bath hotter than the fermion masses, so that the fermions themselves provide the friction.
Only a fraction of the bath energy $\rho_R$ participates in the oscillation.
For the two Dirac fermions, $\Delta E_p=y^2f^2/4E_p$ gives $\mathcal E\le y^4f^4/16\pi^2$, saturated for $T\gg\bar m$, so that
\be
\frac{\mathcal E}{\rho_R}\le\frac{30}{16\pi^4g_*}\left(\frac{yf}{T}\right)^4.
\ee
The masses weakly modulate the energies of only two species, while expansion cools the entire bath.
A stationary bath needs $4H\rho_R=\med{W}\le\Omega\,\mathcal E/2$, that is
\be
\frac{\med{W}}{4H\rho_R}\le\frac{15}{64\pi^4g_*}\,\frac{\Omega}{H}\left(\frac{yf}{T}\right)^4.
\label{eq:Hbound}
\ee
Light fermions, $yf<T$, therefore need many oscillations per Hubble time $\Omega/H\gtrsim416\,g_*$ even for the optimal relaxation rate.
Faster rotation helps only until it outruns collisions: for $\Omega\gg\gamma$ the heating saturates at $\med{W}\approx\gamma\,\mathcal E$.
Fig.~\ref{fig:WLIheating} shows that this maximal heating is below what is needed by all the quartic solutions of~\cite{1805.07186}.
For example, the solution with $Q_*=0.1$ and thermal inflaton modes has $T/H\approx82$, so that $T\le f$ implies $\Omega/H\lesssim610$,
while eq.\eq{Hbound} needs $\Omega/H\gtrsim5200$ even at $yf=T$: relaxation supplies at most $12\%$ of the needed heating.
The deficit is larger for vacuum inflaton modes, for lighter fermions, and for the anomaly-free model of section~\ref{sec:fix},
where the lightness condition and the thermal mass reduce it to about $1\ran3\%$.
The proposed solutions therefore cannot sustain their bath through the assumed quasi-static occupation response.

\medskip

Covering the full parameter space of the model needs lengthy arguments.\footnote{We outline them.
For light fermions, $yf=T$, eq.\eq{Hbound} requires $\Omega/H\ge416\,g_*$.
With energy balance, eq.\eq{deltadot}, this means $T/H\gtrsim600\sqrt{g_*Q}\,f/T$, while the observed amplitude, eq.\eq{PR} with $n_*\approx T/H$ and $G\approx1$, gives $T/H\approx150\,Q^{1/3}$ for $Q\ll1$ and $g_*=18$.
For $T\le f$, $Q\le0.2$ and $T/H\ge1$, the Bose occupation in eq.\eq{PR} gives $(1+R)H/T\le3.1$, so that, for $g_*=18$ and $G=1$, the amplitude requires $T/H\le0.67$: a contradiction, whatever the relaxation rate.
A rate-dependent check leads to the same conclusion.
For $\gamma=\alpha T$, eq.\eq{deltadot} and lightness $T/f\gtrsim y/\sqrt{1-{y'}^2/8}$ imply that $\Omega<\gamma$ holds only for $Q>(2\pi^2g_*/45\alpha^2)\,y^2/(1-{y'}^2/8)$.
Inserting $Q=C_TT/3H$ the Yukawa coupling cancels, and this becomes $T/H>(8\pi^2g_*/45\alpha^2)\,{y'}^2(1-0.34\ln{y'})/(1-{y'}^2/8)\gtrsim10^3g_*$ for the decay-like rate $\alpha={y'}^2/16\pi$, minimised over ${y'}$, while the observed amplitude of perturbations implies $T/H\sim10^2$ for $Q\lesssim1$.}
A numerical example, listed in the first column of table~\ref{tab:examples}, shows how much parameters need to be stretched to make it work:
it needs $T>f$ and an optimal relaxation rate $\gamma=\Omega\approx750\,T$, at the price of a non-perturbative Yukawa coupling $y$ and of an unphysically large ${y'}\sim200$.

\begin{table}[t]
\centering
\begin{tabular}{lccccc}
Model: & \multicolumn{3}{c}{$N=2$ of section~\ref{sec:WLI}} & $N=5$ & $N=\infty$ \\
\cmidrule(lr){2-4}\cmidrule(lr){5-5}\cmidrule(lr){6-6}
Friction mechanism: & \cellcolor{red!8}Relaxation & \multicolumn{3}{>{\columncolor{yellow!20}}c}{Pair production} & \cellcolor{green!10}Ohm\\
\cmidrule(lr){2-2}\cmidrule(lr){3-5}\cmidrule(lr){6-6}
Fermions: & \cellcolor{red!8}light & \cellcolor{yellow!20}light & \cellcolor{yellow!20}heavy & \cellcolor{yellow!20}light & \cellcolor{green!10}light \\
\toprule

Yukawa $y$ & \cellcolor{red!8}4.7 & \cellcolor{yellow!20}0.2 & \cellcolor{yellow!20}1.3 & \cellcolor{yellow!20}0.3 & \cellcolor{green!10}-- \\
Yukawa  ${y'}$ & \cellcolor{red!8}$\color{red!50!black} \sim 200$ & \cellcolor{yellow!20}2 & \cellcolor{yellow!20}-- & \cellcolor{yellow!20}-- & \cellcolor{green!10}-- \\
Degrees of freedom $g_*$ & \cellcolor{red!8}18 & \cellcolor{yellow!20}18 & \cellcolor{yellow!20}18 & \cellcolor{yellow!20}38 & \cellcolor{green!10}470 \\
Friction $Q$ & \cellcolor{red!8}$3~10^{-4}$ & \cellcolor{yellow!20}0.012 & \cellcolor{yellow!20}0.2 & \cellcolor{yellow!20}0.16 & \cellcolor{green!10}$1.1~10^3$ \\
Temperature $T/H$ & \cellcolor{red!8}10 & \cellcolor{yellow!20}34 & \cellcolor{yellow!20}100 & \cellcolor{yellow!20}61 & \cellcolor{green!10}$1.9~10^3$ \\
U(1) breaking scale $f/T$ & \cellcolor{red!8}0.2 & \cellcolor{yellow!20}3.5 & \cellcolor{yellow!20}10 & \cellcolor{yellow!20}1 & \cellcolor{green!10}-- \\
\midrule
Fermion mass $yf/\pi T$ & \cellcolor{red!8}0.30 & \cellcolor{yellow!20}0.22 & \cellcolor{yellow!20}4.1 & \cellcolor{yellow!20}0.095 & \cellcolor{green!10}-- \\
Fermion mass $yf/\Omega$ & \cellcolor{red!8}$0.001$ & \cellcolor{yellow!20}0.1 & \cellcolor{yellow!20}21 & \cellcolor{yellow!20}0.093 & \cellcolor{green!10}-- \\
Modulation $\Omega/T$ & \cellcolor{red!8}750 & \cellcolor{yellow!20}7 & \cellcolor{yellow!20}0.63 & \cellcolor{yellow!20}3.2 & \cellcolor{green!10}-- \\
Modulation $\Omega/H$ & \cellcolor{red!8}7500 & \cellcolor{yellow!20}240 & \cellcolor{yellow!20}63 & \cellcolor{yellow!20}200 & \cellcolor{green!10}-- \\
\end{tabular}
\caption{\em Numerical examples of friction from relaxation of occupations (section~\ref{sec:friction}), from pair production (section~\ref{sec:fixfriction}) and from Ohmic dissipation (section~\ref{sec:WI5d}).
All are reference estimates that impose energy balance and, with thermal inflaton modes, the observed amplitude of perturbations ${\cal P}_R$, within the approximations described in the text.
The relaxation example needs the optimal rate $\gamma=\Omega$, which corresponds to an impossible Yukawa ${y'}\approx200$.
The heavy fermion example needs a nearly maximal efficiency $\epsilon_{\rm dep}\approx0.68$.
The $\mathbb{Z}_5$ ring of section~\ref{sec:ringwarm} has $m_0=yf=\tilde yf$.
The five-dimensional Ohmic example of section~\ref{sec:WI5d} has $\epsilon=10$ and ${\cal R}T=10$, hence a bath rate $\gamma\approx0.4\,T$; its $g_*$ is the effective four-dimensional count of the Kaluza-Klein tower.\label{tab:examples}}
\end{table}

\subsection{Issue 5: ultraviolet behaviour}\label{sec:UV}
The solutions of~\cite{1805.07186} need a large bath coupling, ${y'}\approx2$-$3$, to keep the fermions close to equilibrium, $\Gamma_{\psi_i}>H$.
For the bath of eq.\eq{Lpsi} with a quartic $\lambda_\chi|\chi_i|^4$, the one-loop running at vanishing portal couplings is
\be
(4\pi)^2\frac{d{y'}}{d\ln\bar\mu}=3{y'}^3-\frac34g_X^2{y'},\qquad
(4\pi)^2\frac{d\lambda_\chi}{d\ln\bar\mu}=20\lambda_\chi^2+(8{y'}^2-3g_X^2)\lambda_\chi-4{y'}^4+\frac38g_X^4.
\label{eq:betaS}
\ee
Starting at $\bar\mu=f$, for ${y'}\approx2$ the fermion box $-4{y'}^4$ drives $\lambda_\chi$ negative within less than one $e$-fold above the matching scale for small initial values, and within a few $e$-folds even for $\lambda_\chi=1$; at smaller ${y'}$, large initial values instead grow to non-perturbative values.
Meanwhile ${y'}$ has a Landau pole at $\ln(\bar\mu/f)\approx8\pi^2/3{y'}^2$, a few $e$-folds above $f$: 
this does not invalidate computations at $\bar\mu\sim f$, but leaves little room for the effective theory. 
These issues get even more problematic as the reduced friction of section~\ref{sec:friction} requires larger couplings.
Whether a strongly coupled realisation can overcome these limitations remains an open question.



\subsection{A new friction mechanism: pair production}\label{sec:fixfriction}
The upper bound on friction of eq.\eq{Wbound} assumes a linear relaxational response.
To obtain larger friction we here consider a different process: fermion pair production.
The inflaton usually has a negligible decay width, because the field is roughly time-independent.
In the current model the inflaton provides fermion masses oscillating at $\omega \equiv \Omega/2$,
$m_1=yf\cos(\theta/2)$ and $m_2=yf\sin(\theta/2)$, and effectively behaves as a condensate
of quanta with energy $\omega$ that can decay into fermion pairs.

\subsubsection{Pair production of light fermions}
We assume that fermions always stay light compared to $\omega$.
Each $m_i(t)$ acts as a source for $\bar\psi_i\psi_i$ pair production.
Summing over the two species gives the power $W_{\rm pair}=(yf)^2\omega\,\Im\Pi(\omega)$.
The imaginary part of the retarded  $\bar\psi_i\psi_i$ correlator describes the available phase space:
 for free massless fermions one has $\Im \Pi = \omega^2/8\pi$ at zero temperature, and $\Im\Pi(\omega)=\omega^2\tanh(\omega/4T)/8\pi$ in thermal equilibrium.
Dividing by $\dot\phi^2=f^2\Omega^2$ gives the friction
\be
\Upsilon_{\rm pair}= \frac{W_{\rm pair}}{\dot\phi^2}= \frac{y^2\Omega}{64\pi}\tanh\frac{\Omega}{8T}.
\label{eq:pair}
\ee
For $\Omega\ll T$ the thermal factor suppresses production by $\Omega/8T$.
The heating relative to that needed to sustain the bath is
\be
\frac{W_{\rm pair}}{4H\rho_R}
=\frac{30}{256\pi^3g_*}\frac TH
 \left(\frac{yf}{T}\right)^2\left(\frac\Omega T\right)^3
 \tanh\frac\Omega{8T}.
\label{eq:pairheating}
\ee
Thus a fast rotation can sustain the bath.
Self-consistent solutions $Q_{\rm pair}=Q$ exist,
imposing both energy balance and the observed amplitude, eq.\eq{PR} with $G=1$, $g_*=18$ and thermal inflaton occupation.
An example 
is listed in table~\ref{tab:examples}.
Here $yf\approx0.2\,\omega$, so the weak-source approximation is acceptable.
Treating the thermal mass $m_T\approx0.7\,T$ as a Dirac mass gives the rough kinematic estimate $(1-4m_T^2/\omega^2)^{3/2}\approx0.8$; the actual thermal correction requires the chirally symmetric fermion spectral functions.
The centrifugal correction $\Omega^2/8\lambda_1f^2\approx0.5/\lambda_1$ is small for moderate quartics.
The bath coupling ${y'}$ is not needed for the friction;
$y'$ is needed to thermalise the produced pairs faster than $H$, which for $T/H\approx30$ requires ${y'}\approx2$.
All couplings can be made small enough to avoid the UV issue of section~\ref{sec:UV}. 

\subsubsection{Burst pair production of heavy fermions}
If instead fermions get heavier than $\Omega$, they are pair produced in bursts when each mass crosses zero, as in preheating~\cite{hep-ph/0003018,hep-ph/9812289}.\footnote{A related non-thermal scalar-production mechanism was studied in~\cite{2507.13156}.}
Near a zero the mass varies linearly in time, $m_i\approx\dot m_i\,t$.
Around the zero $\omega_p=\sqrt{p^2+m_i^2}$ is minimal and  changes by order one within a time $p/|\dot m_i|$: the evolution is non-adiabatic if this is shorter than $1/p$, i.e.\ for $p^2\lesssim|\dot m_i|$.
The probability that an initially empty mode ends up occupied by a particle--antiparticle pair is given by the Landau--Zener formula
\be
 P_p=e^{-\pi p^2/p_*^2},\qquad p_*^2=|\dot m_i|=yf\Omega/2,
\ee
so that pairs are produced only with momenta $p\lesssim p_*$.
Counting particles and antiparticles of a Dirac fermion gives $n_{\rm burst}=p_*^3/2\pi^3$.
The two species together cross zero at rate $\Omega/\pi$.
If each produced particle deposits mean energy $E_{\rm dep}$ before the mass falls back, the estimated friction is
\be
\Upsilon_{\rm burst}\sim\epsilon_{\rm dep}
 \frac{y^{5/2}\sqrt{f\Omega}}{4\sqrt2\,\pi^4},
\qquad \epsilon_{\rm dep}\equiv\frac{E_{\rm dep}}{yf}.
\label{eq:burst}
\ee
This expression assumes unsuppressed production and empty incoming modes; thermal gaps and occupied modes reduce or modify it.
A maximal $\epsilon_{\rm dep}=1$ arises if fermions survive while their mass grows and decay before the next zero.
One gets $\epsilon_{\rm dep}\sim\min(1,m_{\rm dec}/yf)$ for a fermion with decay rate 
 $\Gamma_\psi=\beta|m|$ with $\beta\sim{y'}^2/(4\pi)^2$, where $m_{\rm dec}\sim p_*\sqrt{2/\beta}$ is the typical mass at decay.
A bath coupling $y'$ that is too large makes the fermions decay too early, while a coupling that is too small lets them survive until the next zero.

For the example of table~\ref{tab:examples}, eq.\eq{burst} gives $Q_{\rm burst}\sim0.29\epsilon_{\rm dep}$ before thermal suppression: energy balance needs an effective efficiency about $0.68$.
This is close to the maximal efficiency $\approx0.69$ reached for decays before the next zero, while Pauli blocking reduces production by a factor $\approx0.5$ for $p_*\approx2T$.
The example therefore falls short by a factor of a few. 

Furthermore, the large Yukawa $y$ implies that fermions heavier than $\pi T$ generate an unprotected
one-loop correction $-\sum_im_i^4\ln(m_i^2/\bar\mu^2)/(4\pi)^2$ to the inflaton potential, equivalent to a
RGE running 
$\sfrac{d\lambda_5}{d\ln\bar\mu}=-\sfrac{y^4}{64\pi^2}+\cdots$  of the undesired tree-level quartic $\lambda_5$ in eq.\eq{V0}.
This exceeds the bound needed to avoid trapping the inflaton (section~\ref{sec:ringwarm}),
so that $\lambda_5$ must cancel it to a few per cent at $\bar\mu\sim yf$.
The models of section~\ref{sec:ring} achieve better protection.


\medskip

Eq.\eq{PR}, with $G(Q)$ computed for $\Upsilon\propto T$, does not apply to this new friction mechanism, 
because inflaton perturbations feel a friction and a noise different from those of the background.
Perturbations probe the response around the driving frequency.
Their friction is the derivative of the drag with respect to the velocity.
Their noise is the spectrum of the mass operators at the sidebands $\omega\pm\nu$ of the driving frequencies, with $\nu\ll\omega$ the frequency of the perturbation~\cite{1208.3399,2102.09913,2603.11629}.
For example, $\Upsilon_{\rm pair}\propto\dot\phi^2$ for $\Omega\ll T$ gives a drag $\propto\dot\phi^3$, so that perturbations feel the friction $3\Upsilon_{\rm pair}$.
In the burst regime, the random production of particles and their later decay add further noise, as in trapped inflation~\cite{0902.1006}.
Testing these examples against observations therefore needs a dedicated computation of perturbations, with friction and noise obtained from the correlators of the interaction operators as in~\cite{2505.07943}.

\subsection{Issue 6: inflaton potential and occupation}\label{sec:angle}
The inflaton field space is a circle of length $2\pi f$.
So the inflaton potential should be periodic.
A periodic potential could drive natural inflation with decay constant $ \sim f$.
For a single cosine with friction $\Upsilon\propto T$, CMB data still need $f\gtrsim0.3\,\bp$~\cite{2212.04482}.
Light fermions then need a small Yukawa $y\lesssim T/f$, so the friction $C_T\propto y^2$ becomes too small: combining energy balance with the observed amplitude of perturbations excludes weakly dissipative warm solutions with this potential.\footnote{Using eq.s\eq{UpsWLI} and\eq{PR}, energy balance gives $(T/H)^3=45C_TG(Q)(1+R)/(24\pi^4g_*{\cal P}_R)$, so that lightness implies
\be
\frac{T}{H}\lesssim B\,G(Q)(1+R),\qquad \hbox{where}\qquad  B=\frac{45}{32\pi^4g_*{\cal P}_R{y'}^2(1-0.34\ln {y'})}\left(\frac Hf\right)^2\approx10^5\left(\frac Hf\right)^2
\label{eq:coldbound}
\ee
for $g_*=18$ and ${y'}=2$, tiny for $f\sim\bp$, while $1+R\lesssim1+2(1+\pi Q)T/H$ for vacuum or thermal inflaton occupation.}

The inflationary potentials considered in~\cite{1604.08838,1805.07186,2012.03988} are non-periodic up to inflaton values $\phi\gtrsim\bp$ much larger than $f$. 
So $\theta$ winds $\Delta\phi/2\pi f\sim10^{3\ran4}$ times for the quoted values of $f$.
A non-periodic potential requires extending the theory with a monodromy~\cite{0811.1989,0803.3085,0808.0706}.
This produces a multi-branched potential (quadratic in the simplest case), at the price of new ingredients that the model does not contain.

This issue will be addressed in section~\ref{sec:ring}.
The ``Warm Little Inflaton'' model discussed in the current section follows the logic of little Higgs models only in part, and its problems arise in the parts where it departs from it.
In the next section we present simple warm inflation models that follow that logic in full.
We do not call our models ``little'' because  ``little Higgs'' models were proposed to address the ``little hierarchy problem'' of the electroweak scale~\cite{hep-ph/0007265}, 
which has no analogue in inflation, so the word ``little'' is out of context here.

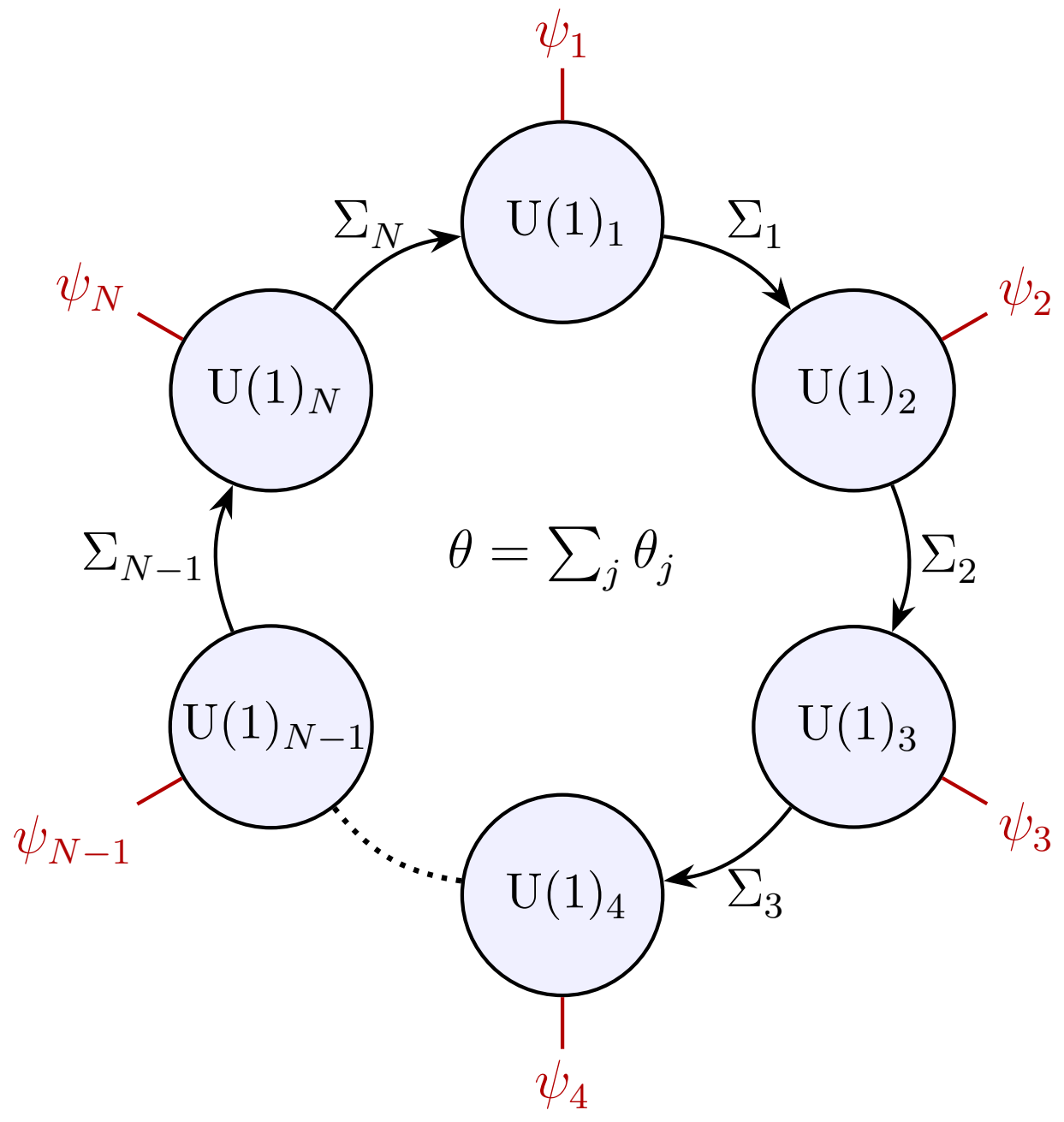
\begin{figure}[t]
\centering
\begin{tikzpicture}[site/.style={circle,draw,thick,fill=blue!6,minimum size=15.5mm,inner sep=0pt},>={Stealth[length=2.5mm]}]
\def\R{2.6}
\foreach \i/\lab in {1/1,2/2,3/3,4/4,5/{N-1},6/N}{
  \pgfmathsetmacro{\ang}{90-60*(\i-1)}
  \node[site] (n\i) at (\ang:\R) {\small$\U(1)_{\lab}$};
  \draw[thick,red!70!black] (n\i.\ang) -- ++(\ang:0.4) node[anchor=\ang+180,inner sep=2pt] {$\psi_{\lab}$};
}
\foreach \i/\j/\lab in {1/2/1,2/3/2,3/4/3,5/6/{N-1},6/1/N}{
  \pgfmathsetmacro{\ang}{60-60*(\i-1)}
  \draw[->,thick] (n\i) to[bend left=22] node[anchor=\ang+180,inner sep=2pt] {$\Sigma_{\lab}$} (n\j);
}
\draw[very thick,dotted] (n4) to[bend left=22] (n5);
\node at (0,0) {$\theta=\sum_j\theta_j$};
\end{tikzpicture}
\caption{\em Circles are the gauge groups $\U(1)_j$, arrows the link scalars $\Sigma_j$ in the bifundamental of $\U(1)_j\otimes\U(1)_{j+1}$, and external legs the Dirac fermions $\psi_j$ charged under $\U(1)_j$.
The $N$ sites can be seen as lattice points of a compact extra dimension, with the link scalars arising from the fifth component of the vector.
The inflaton is the Wilson loop around the ring, the only gauge-invariant phase.}
\label{fig:quiver}
\end{figure}

\section{Ring theories of warm inflation with monodromy}\label{sec:ring}
In this section we propose new models where the symmetry structure of ``little Higgs'' models is fully implemented.
The inspiration is a theory with a fermion and a U(1) vector in one compactified extra dimension, with the inflaton as the Wilson line 
with potential protected by locality.
Section~\ref{sec:5d} will discuss this theory, re-interpret inflaton dissipation as Ohmic conduction along the extra dimension,
and suggest that a monodromy potential arises when the $\U(1)$ is broken by a scalar, as proposed in~\cite{1407.1951} for cold inflation.
We start by presenting four-dimensional models, obtained by replacing the extra dimension with $N$ equi-spaced lattice points~\cite{2001.06518},
and by using the extra freedom to obtain a wider class of models while retaining the good properties of the geometric model.
The case $N=2$ reduces to the model of the previous section.
The subsequent section~\ref{sec:ringwarm} will study if/how these models achieve warm inflation.

\subsection{Warm inflation from fermions on a ring in theory space}\label{sec:Nring}
Following the previous motivation we consider a theory with:
\begin{itemize}
\item gauge group $G=\prod_{j=1}^N \U(1)$ for generic $N$, with indices modulo $N$: $j+N\equiv j$.
\item  $N$ Dirac fermions $\psi_j$ with unit charge under $\U(1)_j$ and neutral under the rest of $G$.
\item $N$ scalars $\Sigma_j$ with charges $(+1,-1)$ under $\U(1)_j\otimes\U(1)_{j+1}$ and neutral under the rest.
\end{itemize}
The fermions are vector-like, so there are no gauge anomalies.
The diagram\footnote{Such diagrams are called `quiver diagrams', the mathematical term for nodes connected by arrows,
or `moose diagrams'  in some physics literature on `deconstruction'. We avoid both terms.}
of fig.~\ref{fig:quiver} illustrates why this structure is special:
a ring of gauge groups with fermions on its sites gives locality in theory space.
This means that for $N\ge3$ the most general renormalisable Yukawa Lagrangian allowed by gauge invariance (for $N=2$ gauge invariance allows further couplings, see section~\ref{sec:N2}) is
\beq
-\Lag_{\rm Yuk}=\sum_{j=1}^N\left[m_j\,\bar\psi_{jL}\psi_{jR}+y_j\,\Sigma_j\,\bar\psi_{jL}\psi_{j+1,R}+\tilde y_j\,\Sigma^*_j\,\bar\psi_{j+1,L}\psi_{jR}\right]+{\rm h.c.}
\label{eq:LYring}
\eeq
The Yukawa couplings $y_j$ and $\tilde y_j$ make a fermion `hop' between neighbouring sites, in the two directions around the ring: we refer to them as forward and reverse `hopping', and to $\tilde y_j=0$ as one-way hopping.
The most general renormalisable scalar potential is\footnote{Replacing U(1) with any other simple group would allow
extra unwelcome invariants, such as $\det\Sigma_j$ for $\SU(N_c)$.}
\beq
V_0=\sum_j\left(-\mu_j^2|\Sigma_j|^2+\lambda_j|\Sigma_j|^4\right)+\sum_{j<k}\lambda_{jk}|\Sigma_j|^2|\Sigma_k|^2+V_{\rm W}.
\label{eq:V0ring}
\eeq
The vacuum expectation value  $\langle\Sigma_j\rangle=f_je^{i\theta_j}$ breaks $G$ to the diagonal U(1)
that couples to all $\psi_j$: the gauge interactions can provide the bath and the widths, without a separate Yukawa bath.

The $\U(1)^N$ gauge transformations remove $N-1$ of the $N$ phases $\theta_j$, leaving the Wilson loop $\theta=\sum_j\theta_j$ as physical scalar.
As in the previous section, this will be the inflaton.
Its kinetic term arises from those of the links, so that the canonical inflaton is $\phi=f_\theta\theta$ with decay constant $f_\theta^{-2}=\sum_jf_j^{-2}/2$, i.e.\ $f_\theta=\sqrt{2/N}\,f$ for equal $f_j=f$.
The vector mass terms $f_j^2(g_jA_j-g_{j+1}A_{j+1})^2$ and the masses of the radial modes depend only on $|\Sigma_j|$.

The `Wilson loop' part of the potential, $V_{\rm W}(\theta)$, is defined as the only part of the potential that depends on $\theta$.
This  term  is present only for small $N$: $V_{\rm W}=\kappa\,\Sigma_1\Sigma_2\Sigma_3+{\rm h.c.}$ for $N=3$, $V_{\rm W}=\lambda_W\Sigma_1\Sigma_2\Sigma_3\Sigma_4+{\rm h.c.}$ for $N=4$ (see section~\ref{sec:N2} for $N=2$), while $V_{\rm W}=0$ for $N\ge5$.
In this case the $\theta$ potential is flat at tree level, and  only arises from fermion loops.

\smallskip

The $N\times N$ fermion mass matrix has the following non vanishing entries
\beq
M_{jj}=m_j,\qquad M_{j,j+1}=y_jf_je^{i\theta_j},\qquad M_{j+1,j}=\tilde y_jf_je^{-i\theta_j}.
\eeq
The loop and thermal corrections depend on $\Tr[(M^\dagger M)^n]$: 
it can be seen as a sum over closed paths of $2n$ steps on the ring, and gauge invariance lets $\theta$ enter only through paths that wind around it.
Therefore the traces are $\theta$ independent for $n<N$ (for $n<N/2$ if both $y_j$ and $\tilde y_j$ are present; phase symmetries cannot forbid $\tilde y_j$ when $m_j\neq0$).
In the simplest case $\tilde y_j=0$ the first $\theta$ dependence is $\Tr[(M^\dagger M)^N]\supset2N\prod_j|m_jy_jf_j|\cos(\theta+\theta_0)$, where the factor $2N$ counts the starting points of the winding paths and their conjugates, and $\theta_0$ collects the coupling phases that cannot be removed by field redefinitions.
Eq.\eq{Fanalytic} then gives
\beq
\delta V_{\rm W} \approx2N c_N\,T^4\prod_j\frac{|m_jy_jf_j|}{\pi^2T^2}\cos(\theta+\theta_0),
\label{eq:deltaVring}
\eeq
suppressed as $T^4(m/\pi T)^{2N}$.
With both hoppings  the suppression reduces to $T^4(m/\pi T)^{N}$ for even $N$ and $T^4(m/\pi T)^{N+1}$ for odd $N$.
The protection of the inflaton potential follows from gauge locality for arbitrary couplings, while warm little inflaton 
in section~\ref{sec:WLI} needs equal couplings.


\subsection{The symmetric ring}\label{sec:symring}
We do not need to impose any extra structure, but simpler explicit results arise assuming an extra
cyclic $\mathbb{Z}_N$ symmetry $j\to j+1$ that rotates the ring.
In the five-dimensional model it corresponds to rotational invariance along the compactified extra dimension.

This $\mathbb{Z}_N$ enforces equal couplings $m_j=m_0$, $y_j=y$, $\tilde y_j=\tilde y$, equal gauge couplings and $f_j=f$; in eq.\eq{V0ring} it sets $\mu_j=\mu$, $\lambda_j=\lambda$, and makes $\lambda_{jk}$ depend only on the distance between the links around the ring.
In the gauge $\theta_j=\theta/N$ the fermion mass matrix is circulant,
\beq
M=m_0\One+yf\,S+\tilde yf\,S^\dagger,\qquad S^N=e^{i\theta}\One,
\eeq
where $S$ is the cyclic shift matrix dressed by the link phases.
In a generic gauge its only non-vanishing entries are $S_{j,j+1}=e^{i\theta_j}$.
The Yukawa $y$ fills the entries above the diagonal (and the corner $M_{N1}$), the Yukawa $\tilde y$ those below it (and $M_{1N}$).
In the gauge $\theta_j=\theta/N$ one has $S=e^{i\theta/N}P$, where $P$ is the cyclic permutation matrix.
The explicit mass matrices for $N=2,3$ are
\beq
M_2=\bpm m_0&yfe^{i\theta/2}+\tilde yfe^{-i\theta/2}\cr yfe^{i\theta/2}+\tilde yfe^{-i\theta/2}&m_0\epm,\quad
M_3=\bpm m_0&yfe^{i\theta/3}&\tilde yfe^{-i\theta/3}\cr \tilde yfe^{-i\theta/3}&m_0&yfe^{i\theta/3}\cr yfe^{i\theta/3}&\tilde yfe^{-i\theta/3}&m_0\epm.
\eeq
$P$ is diagonalised by the discrete Fourier transform, with eigenvalues $e^{2\pi ik/N}$.
So all mass matrices are diagonal in the same basis, independently of $\theta$.
The eigenvalues  are 
\beq
m_k=m_0+yf\,\omega_k+\tilde yf\,\omega_k^*,\qquad \omega_k=e^{i(\theta+2\pi k)/N},\qquad k=1,\ldots,N.
\label{eq:ringmasses}
\eeq
Each squared mass $|m_k|^2$ depends on $\theta$ and oscillates with equal amplitude and period $2\pi N$.
Instead, the sums $\sum_k|m_k|^{2n}$ that affect the potential are independent of $\theta$ for $n<N$ with one-way hopping, and for $n<N/2$ with both hoppings.\footnote{A related model with chiral fermions corresponds to $m_0=\tilde y=0$: the $|m_k|$ are then constant, and $\theta$ becomes an axion, coupled to gauge fields only through anomalies~\cite{2606.02728}.}
This  allows the fermions to respond to the inflaton without affecting its potential.

The $\mathbb{Z}_N$-symmetric model is particularly suitable for warm inflation because this protection is achieved while the individual fermion masses significantly depend on $\theta$.
In general, an eigenvalue depends on $\theta$ only if its eigenvector extends around the ring.
In the symmetric ring the eigenvectors are plane waves in theory space, 
while with generic couplings the ring is a disordered one-dimensional chain, whose modes are Anderson localised,
so that friction gets suppressed: mildly for small rings, by a factor of a few for couplings varying by $\pm50\%$ and $N\lesssim20$, and more strongly at large $N$.

The $m_k$ cross zero (giving burst production) only if $m_0$, $yf$, $\tilde yf$ satisfy one relation:
for real parameters either $|m_0|=f|y+\tilde y|$, or $y=\tilde y$ with $|m_0|<2|y|f$.

\subsection{$N=2$}\label{sec:N2}
For $N=2$ the ring contains the anomaly-free Warm Little Inflaton of section~\ref{sec:fix} as a special choice of couplings,
but gauge invariance alone does not protect it.
The problem is that the link $\Sigma_2^*$ has the same charges as $\Sigma_1$, so gauge invariance cannot distinguish the two links.
Crossed Yukawa couplings such as $\Sigma_2^*\bar\psi_{1L}\psi_{2R}$ are allowed, and the Wilson loop $\Sigma_1\Sigma_2$ has dimension 2, allowing renormalisable terms $\mu^2\Sigma_1\Sigma_2$, $(\Sigma_1\Sigma_2)^2$, $|\Sigma_j|^2\Sigma_1\Sigma_2$ that depend on $\theta$ and also make scalar masses $\theta$ dependent, reintroducing the bosonic cubic term.
Indeed $\U(1)_1\otimes\U(1)_2$ splits into the diagonal $\U(1)_V$, unbroken and acting only on fermions as vector-like fermion number, and the relative $\U(1)_A$, under which $\Sigma_1$ and $\Sigma_2^*$ have equal charge.
The scalar sector therefore coincides with the $\Phi_{1,2}$ formulation of the Warm Little Inflaton, with $\Sigma_1\leftrightarrow\Phi_1$, $\Sigma_2^*\leftrightarrow\Phi_2$ and $\U(1)_A\leftrightarrow\U(1)_X$,
while the vector of $\U(1)_V$ is an extra unbroken vector.

The symmetric ring, whose cyclic $\mathbb{Z}_2$ imposes $y_1=y_2$ and $\tilde y_1=\tilde y_2$, is unsuitable for $N=2$, as $\Tr M^\dagger M$ depends on $\theta$ already at $n=1$ for $y\tilde y\neq0$.

The extra exchange symmetry of the Warm Little Inflaton is a $\mathbb{Z}_2$ different from the cyclic one.
The fermions of section~\ref{sec:fix} are $(\psi_{1L},\psi_{2R})$ and $(\psi_{2L},\psi_{1R})$ in the notation of the ring,
and the off-diagonal entries of the mass matrix are $M_{12}=y_1\Sigma_1+\tilde y_2\Sigma_2^*$ and $M_{21}=y_2\Sigma_2+\tilde y_1\Sigma_1^*$.
The Warm Little Inflaton corresponds to $y_1=y_2=\tilde y_2=-\tilde y_1=y/2$, which gives $M_{12}=yf\cos(\theta/2)$ and $M_{21}=i\,yf\sin(\theta/2)$ in the gauge $\theta_1=\theta_2=\theta/2$,
and to $m_0=0$, as the family-mixing mass is forbidden by its $\mathbb{Z}_4\otimes\mathbb{Z}_4$ symmetry.


In conclusion, for $N=2$ the traces are protected only up to $n=1$, by the exchange symmetry rather than by the ``little Higgs'' structure,
and the crossed couplings must be forbidden by extra discrete symmetries or assumed small.

\subsection{$N>2$}
For $N>2$ the link $\Sigma_k^*$ never has the charges of $\Sigma_j$, so gauge invariance alone forbids crossed and longer-range couplings.

For $N=3$ and one-way hopping, $\tilde y=0$, the $\Tr[(M^\dagger M)^2]$ term cancels and the first thermal $\theta$ dependence is finite, $\sim m^6/T^2$. 
With both hopping directions present, this trace depends on $\theta$ for $N=3$ and $N=4$.
The soft cubic term $\Sigma_1\Sigma_2\Sigma_3$ is an allowed Wilson loop:
with one-way hopping it is generated with a finite coefficient, while with both hoppings its coefficient $\propto m_0y\tilde y(y+\tilde y)$ is logarithmically divergent, and runs.
 
\medskip

For $N=4$ the Wilson loop is a scalar quartic, generated with a running coefficient if both $y$ and $\tilde y$ are present.

For higher $N$ no renormalizable scalar operator depends on $\theta$, 
and a Wilson loop is a non-renormalizable operator
$\prod_j\Sigma_j/\Lambda^{N-4}$, suppressed by an unspecified scale $\Lambda$, possibly the Planck scale.
It does not trap the inflaton if its slope is below the driving force, $f^N/\Lambda^{N-4}\lesssim2\sqrt3(1+Q)f_\theta H\sqrt{\rho_R/Q}$ (see section~\ref{sec:ringwarm}).

\medskip

So far all these models lack the  monodromy potential needed for the inflaton.
We next discuss the extra-dimensional model, that will inspire how to achieve an inflationary potential.

\subsection{$N\to \infty$: the Lord of the Rings}\label{sec:5d}
For $N\to\infty$ the symmetric ring becomes a fermion charged under
a U(1) on a circle: the fermionic analogue of extra-natural inflation~\cite{hep-th/0301218,hep-th/0302034,hep-ph/0205237,2004.13755}.
The ring describes a flat extra dimension, with coordinate $x_5$, compactified on a circle of radius ${\cal R}$ and discretised into $N$ sites with constant
lattice spacing $a=2\pi {\cal R}/N$~\cite{hep-ph/0105239}.
The five-dimensional description holds for $1/{\cal R}\ll T\ll1/a$: many Kaluza-Klein modes are excited, but the lattice is not resolved.
This differs from the regime of eq.\eq{Fanalytic}, where all fermions are lighter than $T$, i.e.\ $T\gg1/a$.
The continuum theory contains:
\begin{itemize}
\item a five-dimensional $\U(1)$ gauge field ${\cal A}_M$ with gauge coupling $g_5$;
\item a five-dimensional Dirac fermion $\Psi$ with unit charge and five-dimensional mass $M$.
\end{itemize}
The theory is vector-like, so there are no anomalies.
The  fifth component ${\cal A}_5$ of the gauge field provides the  links scalars.
The inflaton is its Wilson line around the circle, $\theta=\oint {\cal A}_5\,dx_5$, justifying its name in the four-dimensional model. 
Geometrically, it is the magnetic flux through the circle.
The decay constant of the inflaton becomes $f_\theta=1/2\pi g_4{\cal R}$, where $g_4^2=g_5^2/2\pi {\cal R}$ is the four-dimensional gauge coupling.
The unbroken diagonal $\U(1)$ is the zero mode of ${\cal A}_\mu$.
The Yukawa couplings to the links arise from the covariant derivative $D_5$ in the kinetic term of the five-dimensional fermion.
The $\mathbb{Z}_N$ symmetry corresponds to rotational invariance along the circle.
The fermions fill a Kaluza-Klein tower with masses
\beq\label{eq:KKm}
m_k^2=M^2+\frac{(k+\theta/2\pi)^2}{{\cal R}^2},\qquad k\in\mathbb{Z},
\eeq
reproduced by eq.\eq{ringmasses} of the four-dimensional model for $|k|\ll N$:
the hoppings act as a lattice derivative with $1/a^2=(y-\tilde y)^2f^2-M(y+\tilde y)f$, and the bulk mass is $M=m_0+(y+\tilde y)f$.
The correction $-M(y+\tilde y)f$ is negligible for $M\ll(y-\tilde y)^2f/(y+\tilde y)$, so that $1/a\approx|y-\tilde y|f$.
A second band of light modes would arise only for $M\approx2(y+\tilde y)f$.
The parameter choice that allowed $m_k$ crossing zero corresponds to a massless five-dimensional fermion, $M=0$ up to relabelling the sites.
For $\theta\to\theta+2\pi$ each mode flows into the next one, $k\to k+1$, as in the four-dimensional model.

Any local potential for ${\cal A}_5$ is forbidden by five-dimensional gauge invariance, 
and arises only through fermion world-lines that wind around the circle.
At one loop and high temperature every fermionic Matsubara mode is screened by at least $\pi T$, so that the $\theta$-dependent free energy is suppressed as 
$\exp(-2\pi {\cal R}\sqrt{M^2+\pi^2T^2})$, the continuum analogue of the $(m/\pi T)^{2N}$ suppression of eq.\eq{Fanalytic}.
The zero-temperature Casimir potential $\propto\cos\theta/{\cal R}^4$ is part of this free energy
and is cancelled by the thermal contribution: it reappears only at $T\lesssim1/{\cal R}$.

\medskip

A rolling inflaton is a uniform electric field along the compact dimension~\cite{2508.11206}, 
$E=F_{05}=\dot {\cal A}_5=\dot\theta/2\pi {\cal R}$, constant during slow-roll.
No charges are needed as this electric field is purely inductive: 
it  cannot be written as an electrostatic field $-\partial_5{\cal A}_0$, as ${\cal A}_0$ would not be single-valued around the circle.
In the gauge ${\cal A}_0=0$, the Faraday law $\oint E\,dx_5=\dot\theta$ states that the time-dependent magnetic flux $\theta$ through the circle (an Aharonov-Bohm flux) induces an electromotive force around it, as in a metallic ring threaded by a solenoid with a ramping current.
The energy stored in this field is the inflaton kinetic energy, $\dot\phi^2/2$,
and breaks four-dimensional boost invariance to rotations.
Inflaton friction corresponds to Ohmic dissipation, as discussed later in section~\ref{sec:WI5d}.


\subsection{Monodromy from a superconducting ring}\label{sec:5dmono}
The five-dimensional geometric model suggests how to add the missing monodromy, needed to have an inflationary potential.

We start from a condensed-matter analogy.
A superconductor is an Abelian Higgs model: the electron pairs are described by a charged scalar, the condensate,
with a vacuum expectation value that gives a mass to the photon.
In a superconducting ring threaded by a magnetic flux $\theta$, the phase of this scalar must be single-valued,
so that it winds an integer number $n$ of times around the ring.
The supercurrent is proportional to $\theta-2\pi n$, and the free energy is $F_n(\theta)\propto(\theta-2\pi n)^2$: one parabola for each $n$.
This is a monodromy: $\theta\to\theta+2\pi$ is a symmetry only together with $n\to n+1$,
while at fixed $n$ the energy is not periodic and keeps growing with $\theta$.
The integer $n$ is topological: it can change only where the condensate vanishes,
through a vortex that crosses the width of the ring (a `phase slip').
This costs energy, so that the system can remain on one branch for a long time.

\smallskip

This analogy suggests how to extend the five-dimensional model: by giving a mass to its vector, as in a superconductor.
The resulting model of monodromy inflation was proposed in~\cite{1407.1951,1512.04684}.
We add a five-dimensional scalar ${\cal H}$ with charge $q$, in units of the fermion charge, and with vacuum expectation value $v_5/\sqrt2$.
It plays the role of the condensate.
Writing ${\cal H}=v_5e^{i\sigma}/\sqrt2$, its phase is a compact scalar, $\sigma\equiv\sigma+2\pi$, that shifts under gauge transformations, ${\cal A}_M\to {\cal A}_M+\partial_M\alpha$, $\sigma\to\sigma+q\alpha$.
We assume that its radial mode is heavy, with mass $M_h$, so that only the phase is kept (St\"uckelberg limit).
The kinetic term $|(\partial_M-iq{\cal A}_M){\cal H}|^2$ then becomes a gauge-invariant mass term for the vector,
\beq
\Lag\supset\frac{v_5^2}{2}(\partial_M\sigma-q{\cal A}_M)^2=\frac{M_A^2}{2g_5^2}\left({\cal A}_M-\frac{\partial_M\sigma}{q}\right)^2,\qquad M_A=qg_5v_5,
\label{eq:Stueckelberg}
\eeq
where $M_A$ is the vector mass, given the gauge kinetic term $-F_{MN}F^{MN}/4g_5^2$.
The monodromy arises  because $\sigma$ can wind along the extra dimension, $\sigma(x_5+2\pi {\cal R})=\sigma(x_5)+2\pi n$.
At fixed winding number $n$ the energy is minimised by the uniform winding $\sigma=nx_5/{\cal R}$.
Inserting it and the zero mode ${\cal A}_5=\theta/2\pi {\cal R}$ in eq.\eq{Stueckelberg}, and integrating over the circle, gives the four-dimensional monodromy potential
\beq\label{eq:Vn}
V_n=\frac{M_A^2}{2}f_\theta^2\left(\theta-\frac{2\pi n}{q}\right)^2=\frac{M_A^2}{2}(\phi-\phi_n)^2,\qquad \phi_n=\frac{2\pi n}{q}f_\theta.
\eeq
This simply means that the vector mass $M_A$ is shared by its ${\cal A}_5$ zero mode, the inflaton.\footnote{This mass prevents identifying the U(1) gauge group
with hyper-charge.
Dualising $\sigma$ into a 3-form, the reduction to 4 dimensions gives the coupling of the inflaton to a 4-form~\cite{0811.1989}:
its flux labels the branches $V_n$, and a change of branch is the nucleation of a membrane.}
At fixed condensate, the massive Kaluza-Klein vectors have squared masses ${M_A^2+k^2/{\cal R}^2}$ independent of $\theta$, 
so the bosons do not feel $\theta$ beyond the tree-level monodromy term, while the fermion contribution remains periodic and thermally screened.
With the flattening discussed below, the condensate and therefore $M_A$ depend on $\phi$: the resulting smooth bosonic thermal corrections must be kept smaller than $V_n$ in eq.\eq{Vn}.


The inflaton potential is quadratic only in the limit of a heavy radial mode of the Higgs ${\cal H}$.
Writing ${\cal H}=h\,e^{i\sigma}/\sqrt2$, its kinetic term contains $\frac12h^2(\partial_M\sigma-q{\cal A}_M)^2$, the same term that gives eq.\eq{Stueckelberg} at $h=v_5$.
On a branch $\mu\equiv-\partial_5\sigma+q{\cal A}_5=qg_4(\phi-\phi_n)$ is constant along the circle.
This term therefore gives the radial mode $h$ the extra mass term $\mu^2h^2/2$~\cite{1407.1951}.
For a potential $\lambda(h^2-v_5^2)^2/4$, where the five-dimensional quartic $\lambda$ has dimension of a length, this reduces its vacuum expectation value to $h^2=v_5^2-\mu^2/\lambda$, and integrating it out gives
a flattened quadratic potential
\beq
V_n=\frac12M_A^2(\phi-\phi_n)^2\left[1-\frac{q^2g_4^2(\phi-\phi_n)^2}{M_h^2}\right],\qquad M_h^2=2\lambda v_5^2.
\label{eq:Vflat}
\eeq
An example will be illustrated in fig.\fig{Vn5d}.
A generic $\phi^p$ potential in cold inflation
predicts $n_s=1-(p+2)/2N_e$ and $r=4p/N_e$ at $N_e$ $e$-folds before the end of inflation.
The tensor-to-scalar ratio $r$ is too large for $p\approx2$.
Instead, a flattened quadratic potential with $p \lesssim 2$ is acceptable in warm inflation~\cite{2504.02655}.
For weak dissipation, $Q\ll1$, friction $\Upsilon\propto T$ and thermally occupied inflaton modes, $n_*\approx T/H$ so that $R\approx2T/H \gg 1$ in eq.\eq{PR}, one finds:
\begin{itemize}
\item $n_s=1-(p+2)/3N_e$, which matches the value $n_s=0.9743\pm0.0034$ obtained combining CMB and BAO data~\cite{2503.14452}
for $p\approx2$ and $N_e\sim55$. 
\item $r\approx\sfrac{4p}{N_eR}$: the tensor spectrum is unchanged, while the scalar spectrum of eq.\eq{PR} is enhanced by $R$.
\end{itemize}
The monodromy inflaton potential is applicable as long as no jump in $n$ happens.
The integer $n$ changes through the nucleation of a vortex of $\sigma$, a membrane that appears in four dimensions as a domain wall.
Its thermal rate is exponentially suppressed, as $e^{-S_3/T}$ with\footnote{The membrane has tension $\sigma_w\approx2\pi \ell v_5^2$,
where $\ell={\ln(M_h\min(r_c,{\cal R}))}\sim 1$.
The branches differ by $\Delta V=V_n-V_{n+1}\approx2\pi M_A^2f_\theta(\phi-\phi_n)/q$,
so that a bubble of the lower branch has critical radius $r_c=2\sigma_w/\Delta V$ and energy $S_3=16\pi\sigma_w^3/3\Delta V^2$.
This estimate holds for $r_c$ larger than the wall thickness $1/M_h$, and fails near the end of the branch, where the barrier vanishes.
It also holds for $r_c\ll {\cal R}$, as in the example of section~\ref{sec:WI5d}: the vortex is localised along the circle, and the winding relaxes around the circle in a time $\sim {\cal R}$.
The branches are long lived if $S_3/T$ exceeds the logarithm of the ratio between the nucleation prefactor and $H^4$.
The flattening of eq.\eq{Vflat} also ends the branch unless $v_4^4\gtrsim V_n/\lambda_4$, where $\lambda_4=\lambda/2\pi {\cal R}$ is the four-dimensional quartic.
Both conditions are stronger than $v_4\gtrsim T$, needed to avoid thermal restoration.\label{fot:S3}}
\beq
\frac{S_3}{T}\approx\frac{8\pi }{3}\,\frac{v_4^4 \ell^3}{V_n\,{\cal R}T},\qquad v_4^2=2\pi{\cal R}v_5^2.
\eeq
The branches are therefore long lived up to $V_n \lesssim v_4^4/ {\cal R}T$, up to logarithms $\ell$.
Since $M_A=qg_4v_4$, this needs a scalar with small coupling $qg_4$, while a thermal bath needs a sizeable coupling $g_4$ of the fermions:
the two are compatible only for $q\ll1$, i.e.\ if the fermions have a much larger charge than the scalar.
This conclusion is avoided at finite $N$, where five-dimensional Lorentz invariance no longer ties the inflaton mass to the vector mass.

\subsection{Monodromy at finite $N$: a ring of Josephson junctions}\label{sec:JJ}
The monodromy can be translated into four-dimensional language when the extra dimension is replaced by $N$ lattice sites.
The number of branches becomes finite, equal to $N$~\cite{2001.06518}.
We focus on a renormalisable realisation, where the five-dimensional scalar ${\cal H}$ becomes one Higgs scalar per site:
we add to the model of section~\ref{sec:Nring} $N$ scalars $H_j$ with unit charge under $\U(1)_j$
(the minimal choice that allows a renormalisable hopping along the links) and $\langle H_j\rangle=ve^{i\sigma_j}$.
Each $\U(1)_j$ gets broken, and the theory contains $N$ gauge-invariant phases
\beq
\delta_j=\theta_j+\sigma_{j+1}-\sigma_j
\eeq
rather than just $\theta= \sum_j\delta_j$, which remains coupled to fermions.
For $N\ge5$ the renormalisable operators that depend on $\delta_j$ are the hopping terms of $H_j$ along the links,
\beq
V\supset-\kappa\sum_j\left(H_j^\dagger\Sigma_jH_{j+1}+{\rm h.c.}\right)=-E_{\rm J}\sum_j\cos\delta_j.
\label{eq:VJ}
\eeq
where $E_{\rm J}=2\kappa fv^2$ is the energy of a Josephson junction between two superconductors.
The coefficient $\kappa$ has dimension of mass: it is the lattice version of the kinetic term $|D_5{\cal H}|^2$, which would fix $\kappa=2g^2f$.
The next-to-nearest hopping $-\lambda_\kappa\sum_j(H_j^\dagger\Sigma_j\Sigma_{j+1}H_{j+2}+{\rm h.c.})=-2\lambda_\kappa f^2v^2 \sum_j\cos(\delta_j+\delta_{j+1})$
is also allowed.
This dimensionless coupling would modify the inflaton mass and the end of the branches.
Both couplings break the shift symmetry of the phases $\delta_j$, so that $\lambda_\kappa$ is generated only through $\kappa$, at order $\kappa^2$:
we assume it negligible, which is technically natural.

At fixed $\theta$ the energy is minimised by equal phase differences, $\delta_j=(\theta-2\pi n)/N$, giving the $N$ branches with monodromy potentials
\beq
V_n=-NE_{\rm J}\cos\frac{\theta-2\pi n}{N}=-NE_{\rm J}\cos\frac{\phi-\phi_n}{F},\qquad n=0,\ldots,N-1,
\label{eq:Vnring}
\eeq
where $\phi=f_\theta\theta$ is the canonical inflaton.
The potential is trigonometric with decay constant $F=Nf_\theta=\sqrt{2N}f$, i.e.\ with a period $2\pi F$ longer by a factor $N$,
and can be approximated as quadratic only 
around the minimum of each branch, where the inflaton mass is $m_\phi^2=E_{\rm J}/2f^2=\kappa v^2/f$.
The branch $n$ is a local minimum with respect to the other phases for $|\theta-2\pi n|<N\pi/2$,
where the supercurrent $\partial V_n/\partial\theta$ reaches the critical current of the junctions.
On one branch $\theta$ therefore winds up to $N/4$ times, and the inflaton travels up to $\Delta\phi=\pi F/2$.
This has three consequences.

In the continuum $\kappa=2g^2f$ implies that the inflaton and the vectors acquire the same mass.
At finite $N$   five-dimensional Lorentz invariance does not relate the two masses~\cite{2001.06518}: a small $\kappa$ gives $m_\phi\ll M_A$ while keeping $v\gtrsim T$.
Therefore the scalars need no small charge, their vacuum expectation value is not thermally restored, and the wall tension is not tied to the inflaton mass.
Changes of $n$ are exponentially suppressed\footnote{The defect that changes $n$ is a domain wall across which one $\delta_j$ slips by $2\pi$, a sine-Gordon kink with tension $\sigma_w\approx8\sqrt{2E_{\rm J}}\,\tilde f$, where $\tilde f^2=f^2v (v^2+4f^2)^{-1/2}$ is the coefficient of $(\partial\delta_j)^2$ after integrating out the vectors.
A slip localised on a single link provides an estimate of the action: letting the neighbouring phases and the link radius relax lowers it by a factor of order one, which the large margin below can absorb.
Since $\Delta V=V_n-V_{n+1}\le2\pi E_{\rm J}$, thermal nucleation is suppressed by
${S_3}/{T}={16\pi\sigma_w^3}/{3\,\Delta V^2\,T}\gtrsim4~10^2\,{\tilde f^3}/{f\,m_\phi T}$,
which is large because $m_\phi\ll T\lesssim f,v$.}
until the end of the branch is approached, where the barrier disappears.

The potential is protected: $\kappa$ is a soft term and, for negligible $\lambda_\kappa$, its only source. 
So at linear order in $\kappa$ loop and thermal corrections only renormalise $E_J$, without changing the form of eq.\eq{VJ}.
The vector masses $M_k^2=2g^2v^2+8g^2f^2\sin^2(\pi k/N)$ do not depend on $\theta$.
For frozen radial modes, whose shifts are of order $\kappa$, the only bosons that feel $\theta$ are the phases $\delta_j$ themselves, as light as the inflaton: their cubic thermal term is smaller than the tree-level potential by a factor $\sim Tm_\phi/f^2$.
The $N-1$ non-uniform combinations of the $\delta_j$ have a Hessian $\propto E_{\rm J}\cos\delta_j$ too, and are therefore light spectators with masses $\sim m_\phi$:
they do not couple to the fermions, and their fluctuations must not leave isocurvature perturbations after inflation.

\medskip

Large-field inflation on one branch needs $F=\sqrt{2N}f$ comparable to the Planck mass, i.e.\ many sites if $f\ll M_{\rm Pl}$.
As usual in large-field inflation, the inflaton travels more than $M_{\rm Pl}$, possibly clashing with conjectures about quantum gravity.
$N$ sites add $\sim10N$ degrees of freedom, and the bound $\Lambda\lesssim M_{\rm Pl}/\sqrt{N_{\rm species}}$ on the cut-off of gravity, 
combined with $f\lesssim\Lambda$, limits $F\lesssim M_{\rm Pl}$.
More simply, large $N$ effectively reopens the extra dimension: the cut-off cannot exceed the Planck scale of the five-dimensional theory,
$\Lambda^3\lesssim M_{\rm Pl}^2/2\pi {\cal R}$.
Warm inflation reduces the super-Planckian requirement to $F\gtrsim M_{\rm Pl}/\sqrt{1+Q}$ if  $Q$ is large.

\medskip


\section{Warm inflation in the new ring theories}\label{sec:ringwarm}
We here generalise the computations of section~\ref{sec:WLI} exploring if/how
the new theories discussed in the previous section can sustain warm inflation.

\subsection{Models at finite $N$ with unspecified inflaton potential}
We first consider the symmetric ring of section~\ref{sec:symring}, focusing on how the new effects depend on $N$, 
with results that do not yet depend on the specific monodromy potential.

Concerning the corrections to the potential,
quantum corrections  at $T=0$ are proportional to $\sum_km_k^4\ln m_k^2$, which  is not polynomial in $M^\dagger M$.
This leads to a $\theta$ dependence: suppressed exponentially in $N$ if no eigenvalue crosses zero, and as a power of $N$ otherwise.
So  $N=3$ is useful with one-way hopping and a specified soft term; 
$N=5$ is the smallest ring that forbids a renormalisable Wilson-loop potential and protects both quadratic and quartic traces even with both hoppings.

The kinetic terms of the links give $\Lag_{\rm kin}=f^2(\partial_\mu\theta)^2/N$, so that the canonical inflaton is $\phi=f_\theta\theta$ with $f_\theta=\sqrt{2/N}\,f$,
with period $2\pi f_\theta$.
With one-way hopping the squared fermion masses are modulated at $\Omega=\dot\theta/N$ (rather than $\Omega=\dot\theta$ as in section~\ref{sec:WLI}), and both hoppings add $2\Omega$.
If all particles are light enough to be in the bath the number of degrees of freedom is $g_*=Ng_1$, 
where $g_1$ counts the degrees of freedom per site. 
Eq.\eq{deltadot} generalizes into
\beq
\frac{\Omega}{T}=\frac{T}{f}\sqrt{\frac{\pi^2g_1}{45\,Q}}.
\label{eq:OmRing}
\eeq
As in section~\ref{sec:friction}, the friction mechanism of~\cite{1604.08838,1805.07186} cannot sustain the bath, because the general bound of eq.\eq{Wbound} applies.
Light fermions, $|m_0|+|y|f\lesssim T$, give $\mathcal E\lesssim NT^4/32\pi^2$, so that sustaining the bath needs a fast rotation, $\Omega/H\gtrsim830\,g_1$, which implies a too low ${\cal P}_R$ for the reference solutions of section~\ref{sec:friction}, with thermal inflaton modes and $Q\lesssim1$.
The factors of $N$ cancel: more fermions are heated, but more fermions must be kept in the bath.
With both hoppings $\mathcal E$ can be twice as large, but this does not change the conclusion.

\medskip

We therefore consider friction from fermion pair production, as in section~\ref{sec:fixfriction}.
Each fermion has a constant mass $m_0$ plus the terms $yf$ and $\tilde yf$, whose phases rotate at $\pm\Omega$.
For $|m_k|\ll\Omega$ the estimate of section~\ref{sec:fixfriction} becomes
\beq
\Upsilon_{\rm pair}\approx\frac{(y^2+\tilde y^2)\,\Omega}{16\pi}\tanh\frac{\Omega}{4T},
\eeq
again independent of $N$ at fixed $\Omega$.
For one-way hopping this is the initial production rate: the rotating mass phase also pumps chirality, and the rate describes a stationary friction only if chirality-flipping processes relax the asymmetry fast enough.\footnote{In the limit $m_0=0$ the masses $|m_k|=yf$ are constant and only their phase rotates. This acts as an axial chemical potential $\Omega/2$, so that pairs are produced only until a chiral asymmetry builds up: a stationary heating needs $m_0$, or anomalous processes, to relax it.}
For $y=\tilde y$ the masses are real, and no chiral asymmetry builds up.\footnote{The phase of $\det M$ winds as the curve $m_0+yfe^{ix}+\tilde yfe^{-ix}$, an ellipse centred at $m_0$, winds around the origin:
this needs $|y|\neq|\tilde y|$ and, for real parameters, $|m_0|<|y+\tilde y|f$.
If the fermions are also charged under a non-Abelian group $G$, the inflaton then acquires a $\theta\,G\tilde G$ coupling with unit anomaly coefficient, and sphalerons relax the pumped chirality.
This is minimal warm inflation~\cite{1910.07525}: the ring adds neither an enhanced anomaly nor a protection, and the issues discussed in~\cite{2611.09893} remain.}

Production in bursts needs fermions that become light.
Each mass varies at a rate $|\dot m|\sim yf\,\Omega$, so that bursts are suppressed as $e^{-\pi m_{\rm min}^2/|\dot m|}$,
where $m_{\rm min}$ is the minimum of $|m_k(t)|$ along its periodic evolution, 
and $\dot m$ is evaluate at that time.
For the typical $m_{\rm min}\sim yf$ bursts are suppressed as $e^{-\pi Nyf/\dot\theta}$.
This suppression can be avoided choosing the parameters discussed in section~\ref{sec:ring}  such that the masses cross zero:
for $y=\tilde y$ and $|m_0|<2yf$ they do with $|\dot m|=\Omega\sqrt{4y^2f^2-m_0^2}$, while for $|m_0|=2yf$ they only touch zero.
In this unusual case  production remains unsuppressed for momenta $p\sim(yf\,\Omega^2)^{1/3}$.

\medskip

Concerning thermalisation, no dedicated Yukawa is needed as
the model contains one massless diagonal vector, coupled to each $\psi_j$ as $g/\sqrt N$.
It thermalises the fermions with a rate $\sim g^4T/N$, which decreases with $N$.
The other $N-1$ vectors acquire masses $M_k=2\sqrt2\,gf|\sin(\pi k/N)|$, the lightest being $M_1\approx2\sqrt2\pi gf/N$,
so that at large $N$ part of the tower is populated even for $gf\gg T$.
The modes with $M_k\lesssim T$ thermalise the fermions with a rate $\sim g^4T\ln(1/g)$, while heavier modes give contact interactions suppressed by their masses.

Using eq.\eq{Fanalytic} we check if corrections to the inflaton potential are small enough.
With both hopping directions and equal couplings, 
 the squared fermion masses are
\beq
|m_k|^2=
\big[m_0+(y+\tilde y)f\cos x_k\big]^2+(y-\tilde y)^2f^2\sin^2x_k,
\eeq
with $x_k=(\theta+2\pi k)/N$.
The first possible $\theta$-dependent term in the fermion traces, whose coefficient can vanish for special parameters, is
\beq
\Tr[(M^\dagger M)^{\lceil N/2\rceil}]\supset N\,(y\tilde yf^2)^{\lfloor N/2\rfloor}\cos\theta\times
\begin{cases}
2 & \hbox{for even $N$},\\
(N+1)\,m_0(y+\tilde y)f & \hbox{for odd $N$}.
\end{cases}
\eeq
The correction to the angular potential has the form $\delta V=A\cos\theta$ with $A\sim T^4(m/\pi T)^N$ at large $N$, with both hoppings.
Such a correction does not trap the inflaton if its slope is smaller than the driving force
$|V'|=3H(1+Q)|\dot\phi|$, i.e.\ for $A\lesssim2\sqrt3\,(1+Q)f_\theta H\sqrt{\rho_R/Q}$.

\smallskip

As an example, we take $N=5$, $f=T$, $m_0=yf=\tilde yf=0.3\,T$ and $g_*=Ng_1\approx38$, where $g_1\approx7.5$ counts the light degrees of freedom of each site.
Pair production sustains the bath with $Q\approx0.16$, and the amplitude of perturbations with thermal inflaton modes gives $T/H\approx61$ ($\mathbb{Z}_5$ column of table~\ref{tab:examples}), with the caveats of section~\ref{sec:fixfriction} on applying eq.\eq{PR} to pair production.
The threshold factor $(1-4|m_k|^2/\Omega^2)^{3/2}$ reduces production from the heaviest fermions by about $0.6$.
More importantly, the links rotate at $\Omega\approx3.2\,T$, faster than their radial mode $\sqrt{2\lambda}f$ unless $\lambda$ is well above 5:
otherwise the centrifugal displacement of section~\ref{sec:rotation} reappears.
Energy balance, eq.\eq{OmRing}, gives $\Omega f\sim T^2/\sqrt Q$, so that a strongly coupled radial mode is a generic requirement, rather than a feature of this example.
The condition on $A$ to avoid trapping is largely satisfied.
A complete test would need to evolve the fermion occupations over many cycles, including Pauli blocking and the response of the link radius.

For $N\le4$ the angular potential is corrected by renormalisable couplings:
the first $\theta$-dependent trace has $n\le2$, so that the $\theta$ dependence enters the thermal mass $\propto m^2T^2$
or the running quartic $\propto m^4\ln(\bar\mu/T)$, not suppressed by $T$.
The $N=2$ models avoid the problematic terms by imposing extra discrete symmetries and assuming small couplings.
We estimate that higher loops do not introduce extra problems.\footnote{Treating 
$m_j$, $y_j$ and $\tilde y_j$ as spurions under the flavour symmetries of the theory shows that at
any loop order a $\theta$-dependent term must contain the same combination of couplings as the one-loop winding paths.
Higher loops just correct its coefficient.
}

\medskip

The construction needs the link scalars $\Sigma_j$ to keep their vacuum expectation value $f$ at temperature $T$.
It is reduced by the thermal mass $c_\Sigma T^2$, with $c_\Sigma=g^2/2+\lambda/3+(y^2+\tilde y^2)/12$ plus portal contributions,
and increased by the rotation of each link at $\Omega=\dot\theta/N$.
The resulting $f^2(T)=f^2-(c_\Sigma T^2-\Omega^2)/2\lambda$ remains positive for $T/f<\sqrt{2\lambda/c_\Sigma}\le\sqrt6$ even for slow rotation $\Omega^2\ll c_\Sigma T^2$.
The fermions are lighter than $\pi T$, as needed for eq.\eq{Fanalytic}, for $|m_0|+(|y|+|\tilde y|)f\lesssim\pi T$.
Both conditions allow $T\sim f$ with couplings of order one.

\subsection{Models at finite $N$ with monodromy inflaton potential}\label{sec:Nmono}
We next consider the inflationary potential resulting from the monodromy of section~\ref{sec:JJ}.

The fermion sector is unchanged, unless the  scalars $H_j$ have special charges $0$ or $\pm2$ which allow Yukawa couplings.
Along one monodromy branch the phase of each fermion mass in eq.\eq{ringmasses} changes by less than $2\pi/4$.
The fermion spectrum as a whole repeats up to $N/4$ times.
The phases rotate more slowly than the Hubble rate,
\beq
\frac{\dot\theta}{N}\lesssim\frac{\pi}{2N_e}\,H,
\label{eq:OmJJ}
\eeq
since along a branch $\int dt\,|\dot\theta|/N\le\pi/2$.
A shorter faster stage is possible, but a stage in which the masses oscillate at frequency $\Omega$ lasts at most $\pi H/2\Omega$ $e$-folds.
The friction mechanisms of sections~\ref{sec:friction} and~\ref{sec:fixfriction} need $\Omega\gg H$, e.g.\ $\Omega/H\sim10^2$ in table~\ref{tab:examples},
and therefore cannot sustain the bath on a monodromy branch.

Large $N$ allows one new regime that remains to be explored: most fermions are heavier than $T$  ($yf,\tilde yf\gg T$) while the lightest ones have $m_{\rm min}\lesssim T$.
This regime is worth exploring because:
1) fermion masses change by $\sim yf\gg T$, so that the energies of the thermal fermions change $yf/T$ times faster, and the quasi-static friction grows as $(yf/T)^2$;
2) the inflaton potential remains protected although the expansion of eq.\eq{Fanalytic} no longer applies:
the correction needs a fermion loop around all $N$ sites, which is exponentially suppressed for $N\gg|y-\tilde y|f/\pi T$, i.e.\ when the ring is longer than the thermal wavelength.
In this regime the ring approaches the continuum: we therefore study it in the five-dimensional model, which corresponds to a restricted choice of parameters.

\subsection{Model in 5 dimensions}\label{sec:WI5d}
We study warm inflation in the five-dimensional model of section~\ref{sec:5d}.
The monodromy inflation potential of section~\ref{sec:5dmono} acquires unbounded branches in the continuum,
and is an acceptable inflationary potential.
The question is whether dissipation is sufficient to sustain warm inflation.

As we anticipated, inflaton motion is an electric field $E$ along the extra dimension,
and the charged Kaluza-Klein fermions form a conducting plasma.
The field accelerates particles and collisions relax their distribution.
Normalising the electric field per unit charge one has $dp_5/dt=E$.
This is the continuum limit of the relaxational friction of section~\ref{sec:friction}, 
with the difference that nothing oscillates: the bound of eq.\eq{Wbound} gets replaced by a condition on the field $E$.
Oscillations arise at finite $N$ as a lattice effect: in the four-dimensional models
each fermion energy is periodic in the accumulated phase $\theta$ with period $2\pi N$, so that a constant $\dot\theta$ makes it oscillate at $\Omega=\dot\theta/N$.
Friction is therefore suppressed for $\Omega\gg\gamma$, and the suppression disappears for $N\to\infty$ at fixed $\dot\theta$.\footnote{In condensed-matter language, the momentum along the circle is defined modulo $2\pi/a$, so that the field $E$ makes each mode oscillate with the Bloch frequency $Ea=\dot\theta/N=\Omega$, the modulation frequency of section~\ref{sec:ringwarm}.
The suppression of eq.\eq{lag} for $\Omega\gg\gamma$ is the suppression of the current by Bloch oscillations in a superlattice~\cite{EsakiTsu},
which disappears in the continuum limit $a\to0$.}

\medskip

To perform a five-dimensional computation of friction,
let $\gamma$ be the effective rate at which collisions relax the electric current.
For perturbative collisions it has the form $\gamma=Cg_5^4T^3=C\epsilon^2T$,
where $C$ is a dimensionless numerical coefficient estimated as $C\approx4~10^{-3}$,\footnote{We estimate it 
assuming massless particles with Maxwell-Boltzmann statistics and a drift of opposite sign for opposite charges,
including Bhabha and Compton scatterings and annihilations into two photons,
and taking into account that scatterings among equal charges conserve the drift and do not contribute.
The spin-averaged squared amplitudes in five dimensions, divided by $g_5^4$, are
$2(s^2+u^2)/t^2+2(t^2+u^2)/s^2+6u^2/st-1$ for Bhabha scattering,
$-3(s/u+u/s)-2$ for Compton scattering, and
$\frac92(t/u+u/t)+3$ for annihilation.
Unlike in four dimensions, extra isotropic constant terms appear, and no Coulomb logarithm appears.
Classically, this provides a variational lower bound on $\sigma$.
Pauli blocking, Bose enhancement, screening and a momentum-dependent drift can change this estimate by factors of order one.
}
and $\epsilon \equiv g_5^2T=2\pi {\cal R}\,g_4^2T $ is the dimensionless strength of the five-dimensional gauge interaction at the scale $T$.
If the momentum gained between collisions is small compared to the temperature, $E\ll\gamma T$, the number current obeys Ohm's law, $J=\sigma E$,
and the bath is heated by the Joule power
\beq
W=\int_0^{2\pi{\cal R}}dx_5\,JE=2\pi {\cal R}\,\sigma E^2,\qquad \sigma=\frac{9\zeta(3)}{16\pi^2}\,\frac{T^3}{\gamma}.
\label{eq:W5d}
\eeq
The conductivity $\sigma$ is here computed for $M\ll T$ and for a momentum-independent $\gamma$.
This power is taken from the inflaton: since $E=\dot\theta/2\pi {\cal R}=g_4\dot\phi$, the friction is
\beq
\Upsilon\equiv \frac{W}{\dot\phi^2}=2\pi {\cal R}\,g_4^2\,\sigma=\frac{9\zeta(3)}{16\pi^2}\,\frac{1}{C g_5^2}.
\label{eq:Ups5d}
\eeq
The friction does not depend on the temperature for a bulk mass $M\ll T$, up to a dependence of $C$ on $\epsilon$ due to screening, discussed below.
The same $\Upsilon$ is obtained summing eq.\eq{UpsWLI} over the Kaluza-Klein tower.

\medskip

For $T\gg1/{\cal R}$ the bath is a five-dimensional relativistic gas, with four-dimensional energy density and pressure
\beq
\rho=\frac{6\zeta(5)}{\pi}\,g_{*5}\,{\cal R}T^5,\qquad p=\frac\rho4,\qquad g_{*5}=g_B+\frac{15}{16}g_F=\frac{31}4,
\eeq
counting the four states of the five-dimensional Dirac fermion and the four polarisations of the massive five-dimensional vector, for $M_A\ll T$ (its longitudinal component
might not thermalize for small Higgs charge).
Since only three dimensions expand, the continuity equation is $\dot\rho+\frac{15}4H\rho=\Upsilon\dot\phi^2$, so that the stationary regime has $\dot\phi^2=5\rho/4Q$ rather than $4\rho/3Q$.
Linear conduction, $E=g_4\dot\phi\ll\gamma T$, requires strong dissipation, $Q\gg3/C^2\epsilon^3$, 
having used $\dot\phi^2=5\rho/4Q$ and $Q=\Upsilon/3H$.\footnote{For stronger fields, $E\gtrsim\gamma T$, the response is no longer Ohmic.
The current of the thermal charges is then bounded by their density, $J\le n=7\pi^2T^4/240$, so that energy balance needs $E\gtrsim32\,HT$.
However the charges then gain more than $T$ between collisions and are no longer thermal: this regime needs a dedicated computation.}

\smallskip

\begin{figure}[t]
\centering
\includegraphics[width=0.65\textwidth]{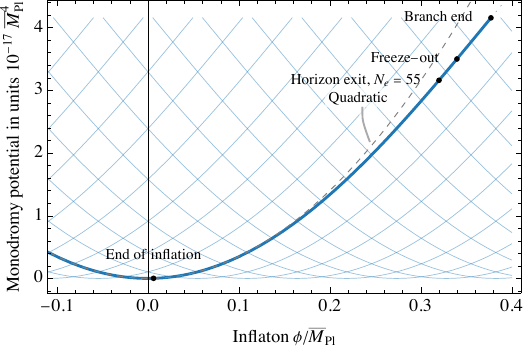}
\caption{\em One branch of the five-dimensional monodromy potential of eq.\eq{Vflat} (thick), and every 10th neighbouring branch (thin),
for the numerical example of section~\ref{sec:WI5d}.
Dots mark the end of inflation, horizon exit at $N_e=55$, freeze-out of thermal perturbations, 
and the end of the branch.
The dashed parabola shows the quadratic approximation.
 \label{fig:Vn5d}}
\end{figure}

The perturbation spectrum is given by eq.\eq{PR}
with $G=1$ since a temperature-independent $\Upsilon$ removes the direct temperature-dependent coupling~\cite{0905.3500},
provided that friction and noise are local, and that radion and phase fluctuations are heavy;
the five-dimensional equation of state of the bath then enters only through metric perturbations, which should give slow-roll corrections.\footnote{The standard
expression for the friction force $\Upsilon\,\delta\dot\phi$ assumes that the force at a point depends only on the inflaton velocity at the same point and time, 
 and that the noise is uncorrelated.
In time the condition is satisfied, as modes near freeze-out $k_{\rm fo}=\sqrt{H\Upsilon}$ evolve at a rate $\sim H\ll\gamma$~\cite{2505.07943}.
In space, locality holds for perturbations on scales longer than
the mean free path $1/\gamma$ of the charges, their thermal wavelength $1/T$, and the radius ${\cal R}$ of the circle around which the current flows.}
For $Q\gg1$ it becomes
\beq
{\cal P}_R\approx
\sqrt{3\pi Q}\,\frac TH\left(\frac{H^2}{2\pi\dot\phi}\right)^2  .
\label{eq:OhmicSpectrum}
\eeq
Inserting $\dot\phi^2=5\rho/4Q$ and $Q=\Upsilon/3H$, the observed ${\cal P}_R$ is reproduced for
\beq
\frac TH\approx\frac{350}{C\epsilon\,({\cal R}T)^{2/3}},\qquad
\hbox{and therefore}\qquad
Q\approx\frac{8}{C^2\epsilon^2({\cal R}T)^{2/3}}.
\label{eq:OhmicWindow}
\eeq
As a numerical example, we assume ${\cal R}T\approx10$ (in the five-dimensional regime left to be explored by section~\ref{sec:Nmono})
and $\epsilon\approx10$ (so that the five-dimensional theory  is mildly weakly coupled at $T$, as the loop factor is $\epsilon/24\pi^3$: $T$ is two orders of magnitude below the cut-off $\Lambda_5\sim24\pi^3/g_5^2$ of the five-dimensional theory).
This corresponds to (last column of table~\ref{tab:examples}) $T/H\approx1.9~10^3$;
$Q\approx1.1~10^3$;
a rate $\gamma\approx0.4\,T$;
a freeze-out scale of thermal perturbations
$k_{\rm fo}=\sqrt{H\Upsilon}\approx0.07\,\gamma$ small enough for a local description;
a sub-Planckian excursion $|\phi-\phi_n|\sim0.3\,\bp$.
A branch extends up to $qg_4|\phi-\phi_n|\approx M_h/\sqrt6$, 
where the supercurrent is maximal and the uniform winding becomes unstable.
For $M_h\lesssim\Lambda_5$ this excursion needs a small charge of the scalar, as discussed in section~\ref{sec:5dmono}:
fig.\fig{Vn5d} shows a numerical example with $q\approx1.6~10^{-4}$ and radial mass $M_h\approx10\,T$.
This corresponds to a four-dimensional quartic $\lambda/(2\pi {\cal R})\approx0.01$ for ${\cal H}$,
small enough that thermal transitions to a lower branch are negligible throughout
the field range needed for inflation, as $S_3/T\gtrsim200$
according to improved estimates along the lines of footnote~\ref{fot:S3}.

Order unity corrections are expected because
the electric field  $\sfrac{E}{\gamma T}\approx0.6\,\sfrac{({\cal R}T)^{1/3}}{\sqrt\epsilon} \approx0.4$ is only mildly in the linear-conduction regime,
and because the neglected vector thermal Debye mass
$m_D^2=9\zeta(3)g_5^2T^3/4\pi^2$
is $m_D \approx 1.7 T$.
In general ${\cal P}_R$ fixes
$(\sfrac{m_D}{T})(\sfrac{E}{\gamma T})\approx0.3\,({\cal R}T)^{1/3}$,
so that screening or non-linear conduction gives corrections of order one.
Extra computations are needed to achieve a precise prediction.
In particular, screening makes $C$ temperature dependent, with $d\ln\Upsilon/d\ln T=-d\ln C/d\ln\epsilon$, modifying both the amplitude and the tilt.

\medskip

As enough friction seems possible, we compute other inflationary observables in the constant-$C$, linear-conduction approximation.
Inserting the slow-roll solution $\dot\phi=-V'/\Upsilon$ and $T^5\propto V'^2/H$, the spectral index is
\beq
n_s-1\approx\frac{-23\epsilon_V+16\eta_V}{10\,Q},\qquad
\epsilon_V=\frac{\bp^2}{2}\frac{V'^2}{V^2},\qquad
\eta_V=\bp^2\frac{V''}V.
\label{eq:OhmicTilt}
\eeq
A quadratic potential gives $n_s\approx1-7/10(N_e+1)\approx0.988$.
The flattening of the monodromy potential in eq.\eq{Vflat} makes the spectrum redder, down to $n_s\approx0.974$ for $N_e=55$ at the end of the branch, $q^2g_4^2(\phi-\phi_n)^2=M_h^2/6$.
In the numerical example above, horizon exit at $q^2g_4^2(\phi-\phi_n)^2\approx0.12\,M_h^2$ gives $n_s\approx0.978$,
and freeze-out happens about 4 $e$-folds earlier, still within the branch, as shown in fig.~\ref{fig:Vn5d}.
Local stability does not ensure survival against thermal phase slips: the nucleation exponent for this flattened background remains to be checked beyond the thin-wall estimate of section~\ref{sec:5dmono}.
The running of the spectral index, $|dn_s/d\ln k|\lesssim2~10^{-4}$, is far below current bounds.


\medskip

To conclude, we consider the pair-production friction mechanisms of section~\ref{sec:fixfriction}:
\begin{itemize}
\item {\em Perturbative pair production} does not survive in the continuum: it needs masses modulated at a frequency above $2 m_k$,
while the Bloch modulation $\Omega=Ea$ vanishes for $a\to0$.
\item {\em Burst production} needs masses that cross zero, 
while the Kaluza-Klein masses of eq.\eq{KKm} never do, because of the bulk mass $M$.
Only $p_5=(k+\theta/2\pi)/{\cal R}$ crosses zero, with $\dot p_5=E$, so that pair-production is suppressed
by the Landau-Zener factor $e^{-\pi(\vec{p}^2+M^2)/E}$.
In five-dimensional language this is Schwinger pair-production by the electric field $E$~\cite{2508.11206,2608.00158}, 
with exponentially suppressed total rate of pair production $2{\cal R} E^{5/2}e^{-\pi M^2/E}/(2\pi)^3$ if $M^2\gtrsim E$.
Even for $M=0$, production is Pauli blocked and subdominant to the thermal charges unless $E\gtrsim T^2$.
In this regime, Schwinger production can sustain a self-regulated plasma with $T/H\lesssim32$,
assuming a saturated current and an equilibrium particle density; charges produced by the field can exceed this density, relaxing the bound.
Applying the Ohmic fluctuation formula would then suggest excessive perturbations;
a dedicated computation is needed.

\end{itemize}
The pressure $p=\rho/4$ also acts along the compact dimension: the five-dimensional radius must be stabilised by a potential stiffer than the thermal and Casimir pressures, 
so that its thermal shift is small, a condition separate from a radion heavier than $H$.
This is the analogue of the centrifugal force in section~\ref{sec:rotation}.


\section{Conclusions}\label{sec:concl}
Warm inflation needs particles coupled to the inflaton strongly enough to keep a thermal bath hot during inflation,
while their thermal corrections must not spoil the flatness of the inflaton potential.
The warm little inflaton model~\cite{1604.08838} proposed to separate the two effects using two fermions:
their individual masses oscillate depending on the inflaton and provide friction, while the sum of their squared masses, which controls the leading thermal correction, does not.

\smallskip

We found that this model has several problems.
Its gauge symmetry is anomalous, and its common bath gives thermal masses different from those assumed:
both problems are fixed by giving opposite charges and separate baths to the two fermions (sections~\ref{sec:anomaly}--\ref{sec:fix}).
The main problem is that friction was over-estimated (section~\ref{sec:friction}):
the inflaton is an angle that rotates many times per Hubble time, modulating the fermion masses much faster than collisions relax the fermion occupations.
As in a microwave oven driven far above the relaxation rate of water molecules, the occupations cannot follow the modulation and little energy is dissipated.
Even assuming the optimal relaxation rate, the total available power shows that two light fermions store too little energy to keep the whole bath hot.

We proposed new friction mechanisms, based on pair production of fermions by their oscillating masses (section~\ref{sec:fixfriction}):
perturbative production of light fermions, and production in bursts when heavy fermions become massless.
They are not restricted by the bound and can sustain the bath, but need fast oscillations, and their perturbations need a dedicated computation.
Furthermore, the fast rotation pushes the scalars away from their assumed vacuum (section~\ref{sec:rotation}), the needed bath couplings are close to Landau poles (section~\ref{sec:UV}),
and the assumed non-periodic inflationary potentials are not compatible with an angular inflaton (section~\ref{sec:angle}).

\smallskip

We then proposed new anomaly-free models that implement the `little Higgs' logic in full:
a ring of $N$ U(1) gauge groups, with one vector-like fermion on each site and scalars on the links (section~\ref{sec:Nring}).
The inflaton is the gauge-invariant Wilson loop around the ring.
By locality in theory space, corrections to its potential arise only from fermion loops that wind around the whole ring,
so that the potential is protected for $N\ge5$ for arbitrary couplings, while the individual fermion masses depend on the
inflaton (section~\ref{sec:symring}).
The warm little inflaton is a special $N=2$ case, where protection
instead relies on extra symmetries (section~\ref{sec:N2}).
These models still lack an inflationary potential.

\smallskip

The ring is the lattice version of a fermion charged under a U(1) in one compact extra dimension (section~\ref{sec:5d}).
No extra scalars are needed.
In this language the warm little inflaton is a two-site discretisation, plus a special choice of couplings, of an electric field driving a current around a compact dimension:
the inflaton is the magnetic flux through the circle, and its growth induces a constant electric field along it.
This picture, by analogy with condensed matter, suggests how to add a monodromy that generates an inflationary potential.
In a superconducting ring the energy grows quadratically with the magnetic flux through it, on branches labelled by the winding of the condensate,
which can change only through rare phase slips.
Adding a charged Higgs in the extra dimension therefore gives the inflaton a monodromy potential with long branches (section~\ref{sec:5dmono}),
and its lattice version gives $N$ branches with a trigonometric potential (section~\ref{sec:JJ}).
However, along one branch at small $N$ the fermion masses oscillate more slowly than the Hubble rate, 
so that neither relaxation nor pair production provides enough friction (sections~\ref{sec:ringwarm} and~\ref{sec:Nmono}).
The reason is that fermion mass oscillations are lattice artefacts that  disappear in the continuum limit $N\to\infty$,
where simpler physics emerges: one ring to rule them all.

The charged Kaluza-Klein fermions form a plasma, and friction is the familiar Ohmic dissipation of the current driven by the electric field (section~\ref{sec:WI5d}).
The friction does not depend on the temperature, and is large enough to sustain warm inflation in the strong-dissipation regime,
for a five-dimensional gauge interaction mildly weakly coupled at the temperature $T$, and a radius ${\cal R}$ moderately larger than $1/T$.
In the superconducting ring the electric field drives two currents:
a reversible supercurrent, which stores energy and provides the inflaton potential, controlled by the vector mass;
and a dissipative normal current of fermions, which provides the friction, controlled by their conductivity.
A flattened quadratic potential gives a spectral index compatible with data, 
provided that the Higgs charge is much smaller than the fermion charge.
This success sits at the edge of control.

%

{\footnotesize
\paragraph{Acknowledgements.}
AG acknowledges support from the Royal Society, UK, funding reference: NIF\ R1\ 253963.
AS thanks Anthropic for a `Claude Team plan for scientists' discount.
We thank ChatGPT for comments, and Claude, which took an active part in this work.
In the Indian tradition, the learned Ga\d{n}e\'sa agreed to transcribe the Sanskrit epic
\emph{Mah\=abh\=arata}, as the sage Vy\=asa dictated it, on condition that the dictation
never pause. Vy\=asa accepted, with a condition of his own: Ga\d{n}e\'sa must
understand each verse before committing it to writing.

\end{document}